\documentclass[10pt,final,twocolumn,journal,twoside]{IEEEtran}

\newtheorem{Thm}{Theorem}

\newtheorem{Prop}{Proposition}

\usepackage{amsmath}
\usepackage{amssymb}
\usepackage{psfrag}
\usepackage{subfigure}
\usepackage{graphicx}

\newcommand{\argmin}{\mathop{\mathrm{arg\min}}}

\title{Minimum-Cost Multicast over \\ Coded Packet Networks}

\author{Desmond S. Lun,
Niranjan Ratnakar,
Muriel M\'edard,
Ralf Koetter, \\
David R. Karger,
Tracey Ho, Ebad Ahmed, and Fang Zhao
\thanks{This work was supported by 
the National Science Foundation under 
grant nos.\  CCR-0093349, CCR-0325496, and CCR-0325673; 
by the Army Research Office through University of California
subaward no.\  S0176938;
by the Office of Naval Research under 
grant no.\  N00014-05-1-0197; 
and by the Vodafone Foundation.}
\thanks{This paper was presented in part at the International Symposium on
Information Theory and its Applications, Parma, Italy, October 2004; 
in part at IEEE Infocom, Miami, FL, March 2005; 
in part 
at the First Workshop on Network Coding, Theory, and Applications,
Riva del Garda, Italy, April 2005;
and in part at the International Zurich Seminar on Communications,
Zurich, Switzerland, February 2006}
\thanks{D. S. Lun, M. M\'edard, 
E. Ahmed, and F. Zhao
are with the Laboratory for
Information and Decision Systems, Massachusetts Institute of
Technology, Cambridge, MA 02139, USA (e-mail: dslun@mit.edu,
medard@mit.edu, ebad@mit.edu, zhaof@mit.edu).}
\thanks{N. Ratnakar and R. Koetter 
are with the Coordinated Science Laboratory,
University of Illinois at Urbana-Champaign, 
Urbana, IL 61801, USA (e-mail: ratnakar@uiuc.edu, koetter@uiuc.edu).}
\thanks{D. R. Karger is with the Computer Science and
Artificial Intelligence Laboratory, Massachusetts Institute of
Technology, Cambridge, MA 02139, USA (e-mail: karger@mit.edu).}
\thanks{T. Ho is with the Department of Electrical Engineering,
California Institute of Technology, Pasadena, CA 91125, USA
(e-mail: tho@caltech.edu)}}

\begin{document}
\maketitle

\begin{abstract}
We consider the problem of establishing minimum-cost multicast
connections over coded packet networks, i.e.\  packet networks where the
contents of outgoing packets are arbitrary, causal functions of the
contents of received packets.  We consider both wireline and wireless
packet networks as well as both static multicast (where membership of
the multicast group remains constant for the duration of the connection)
and dynamic multicast (where membership of the multicast group changes
in time, with nodes joining and leaving the group).

For static multicast, we reduce the problem to a polynomial-time
solvable optimization problem, and we present decentralized algorithms
for solving it.  These algorithms, when coupled with existing
decentralized schemes for constructing network codes, yield a fully
decentralized approach for achieving minimum-cost multicast.  By
contrast, establishing minimum-cost static 
multicast connections over routed
packet networks is a very difficult problem even using centralized
computation, except in the special cases of unicast and broadcast
connections.

For dynamic multicast, we reduce the problem to a dynamic programming
problem and apply the theory of dynamic programming to suggest how it
may be solved.  
\end{abstract}

\begin{keywords}
Ad hoc networks, communication networks,
distributed algorithms, dynamic multicast groups, 
multicast, network coding, 
network optimization, wireless networks
\end{keywords}

\IEEEpeerreviewmaketitle

\section{Introduction}

A typical node in today's packet networks is capable of two functions:
forwarding (i.e.\  copying an incoming packet onto an outgoing link) and
replicating (i.e.\  copying an incoming packet onto several outgoing
links).  But there is no intrinsic reason why we must assume these are
the only functions ever permitted to nodes and, in application-level
overlay networks and multi-hop wireless networks, for example, allowing
nodes to have a wider variety of functions makes sense.  We therefore
consider packet networks where the contents of outgoing packets are
arbitrary, causal functions of the contents of received packets, and we
call such networks coded packet networks.

Coded packet networks were put forward 
by Ahlswede et al.\  \cite{acl00}, and 
numerous subsequent papers,
e.g.,\  \cite{lyc03, kom03, jsc05, hmk, lmk},
have built upon their work.
These papers, however, all assume the availability of dedicated network
resources, and scant 
attention is paid to the problem of determining the
allocation of network resources to dedicate to a particular connection
or set of connections.
This is the problem we tackle.
More precisely, we aim to find minimum-cost subgraphs
that allow given multicast connections to be established
(with appropriate coding) over coded packet networks.

The analogous problem for routed packet networks is old and
difficult.
It dates to the 1980s and, in the simplest case---that
of static multicast in wireline networks with linear cost---it 
equates to the Steiner
tree problem, which is well-known to be NP-complete \cite{bhj83, wax88}.
The emphasis, therefore, has been on heuristic methods.
These methods include heuristics for the Steiner tree problem on
undirected (e.g.,\  \cite{bhj83, win87, wax88}) 
and directed (e.g.,\  \cite{ram96, ccc99, zok02}) graphs, 
for multicast tree generation in wireless networks 
(e.g.\ \cite{wne02b}),
and for the dynamic or on-line Steiner tree problem 
(e.g.,\  \cite{wax88, imw91, wey93}).
Finding minimum-cost subgraphs in coded packet
networks, however, is much easier and as we shall see, in many cases,
we are able to find optimal
subgraphs in polynomial time using decentralized computation.
Moreover, since coded packet networks are less constrained than routed
ones, the minimum cost for a given connection
is generally less.

In our problem, we take given multicast connections and thus include
unicast and broadcast connections as special cases.
But we do not consider optimizing the subgraph for multiple connections
taking place simultaneously.  One reason for this is that 
coding for multiple connections is a very difficult
problem---one that, in fact, remains currently open with only 
cumbersome bounds on the asymptotic capability of coding \cite{syc03}
and examples that demonstrate the insufficiency of various classes of
linear codes \cite{mek03, ral03, rii04, dfz05}.
An obvious, but sub-optimal, approach to coding is to code for
each connection separately,
which is referred to as superposition coding \cite{yeu95}.
When using superposition coding, finding minimum-cost allocations
for multiple connections means extending the approach for single
connections (namely, the approach taken in this paper)
in a straightforward way that is completely analogous to
the extension that needs to be done for traditional 
routed packet networks,
and this problem of minimum-cost allocations for multiple connections
using superposition coding is addressed in \cite{cxn04}.
An alternative approach to coding that outperforms
superposition coding, but that remains sub-optimal, is discussed in
\cite{lmh04}.

We choose here
to restrict our attention to single connections because
the subgraph selection problem is simpler
and because minimum-cost single
connections are interesting in their own right:  Whenever 
each multicast group has a selfish cost objective, or when the network
sets link weights to meet its objective or enforce certain policies and
each multicast group is subject to a minimum-weight objective, we wish to
set up single multicast connections at minimum cost.

Finally, we mention that a related problem to subgraph selection, that
of throughput maximization, is studied for coded networks
in \cite{llj05, lil05} and that an
alternative formulation of the subgraph selection problem for coded
wireless packet networks is given in \cite{wck05}.

The body of this paper is composed of four sections:
Sections~\ref{sec:wireline} and~\ref{sec:wireless} deal with static
multicast (where membership of the multicast group remains constant for
the duration of the connection) for wireline and wireless packet
networks, respectively; Section~\ref{sec:comparison} gives a comparison
of the proposed techniques for static multicast with techniques in
routed packet networks; and Section~\ref{sec:dynamic} deals with dynamic
multicast (where membership of the multicast group changes in time, with
nodes joining and leaving the group).  We conclude in
Section~\ref{sec:conclusion} and, in so doing, we give a sampling of the
avenues for future investigation that our work opens up.  

\section{Wireline packet networks}
\label{sec:wireline}

We represent the network with a directed graph
$\mathcal{G}=(\mathcal{N},\mathcal{A})$, where $\mathcal{N}$ is
the set of nodes and $\mathcal{A}$ is the set of arcs.  Each arc $(i,j)$
represents a lossless point-to-point link from node $i$ to node $j$.  
We denote by $z_{ij}$ the rate at which coded packets are injected into
arc $(i,j)$.
The rate vector $z$, consisting of $z_{ij}$, 
$(i,j) \in \mathcal{A}$, is called a subgraph, and we assume that it
must lie within a constraint set $Z$ for, if
not, the packet queues associated with one or more arcs becomes 
unstable.  We reasonably assume that $Z$ is 
a convex subset of the positive orthant containing the origin.
We associate with the network a cost function $f$ (reflecting, for example,
the average latency or energy consumption) that maps valid rate
vectors to real numbers and that we seek to minimize.

Suppose we have a source node $s$ wishing to transmit packets at a
positive, real rate $R$ to a non-empty set of sink nodes $T$.
Consider the following optimization problem:
\begin{equation}
\begin{split}
& \begin{aligned}
\text{minimize }   & f(z) \\
\text{subject to } 
& z \in Z, \\
& z_{ij} \ge x_{ij}^{(t)} \ge 0,
  \qquad \text{$\forall$ $(i,j) \in A$, $t \in T$} , 
\end{aligned} \\
&\; \sum_{\{j | (i,j) \in A\}} x_{ij}^{(t)}
  - \sum_{\{j | (j,i) \in A\}} x_{ji}^{(t)} 
= \sigma_i^{(t)}, \\
& \qquad \qquad \qquad \qquad \qquad \qquad
\qquad \text{$\forall$ $i \in N$, $t \in T$} , 
\end{split}
\label{eqn:1}
\end{equation}
where
\[
\sigma_i^{(t)} =
\begin{cases}
R & \text{if $i = s$}, \\
-R & \text{if $i = t$}, \\
0 & \text{otherwise}.
\end{cases} \\
\]

\begin{Thm} 
The vector $z$ is part of a feasible solution for the optimization
problem (\ref{eqn:1}) if and only if there exists a network code that
sets up a multicast connection in the wireline network represented by
graph $\mathcal{G}$ at rate arbitrarily close to $R$ from source $s$ to
sinks in the set $T$ and that injects packets at rate arbitrarily close
to $z_{ij}$ on each arc $(i,j)$.  
\label{thm:4} 
\end{Thm}

\begin{proof}
First suppose that $z$ is part of
a feasible solution for the problem.  Then, for any $t$ in $T$,
we see that the maximum flow
from $s$ to $t$ in the network 
where each arc $(i,j)$ has maximum input rate
$z_{ij}$ is at least $R$.  
So, by Theorem~1 of \cite{acl00}, a coding solution that injects packets
at rate arbitrarily close to $z_{ij}$ on each arc $(i,j)$ exists.
Conversely, suppose that we have a coding solution that injects packets
at rate arbitrarily close to
$z_{ij}$ on each arc $(i,j)$.  Then the maximum input rate of
each arc must be at least $z_{ij}$ and moreover, 
again by Theorem~1 of \cite{acl00}, flows of size $R$ exist
from $s$ to $t$ for each $t$ in $T$.
Therefore the vector $z$ is part of 
a feasible solution for the optimization problem.
\end{proof}

From Theorem~\ref{thm:4}, it follows immediately that
optimization problem (\ref{eqn:1}) finds the optimal cost for
an asymptotically-achievable, rate-$R$  
multicast connection from $s$ to $T$.

\begin{figure}
\centering
\psfrag{#s#}{$s$}
\psfrag{#t_1#}{$t_1$}
\psfrag{#t_2#}{$t_2$}
\psfrag{1}{\small 1}
\psfrag{2}{\small 2}
\psfrag{3}{\small 3}
\subfigure[Each arc is marked with its cost per unit rate.]
{\includegraphics[scale=1.2]{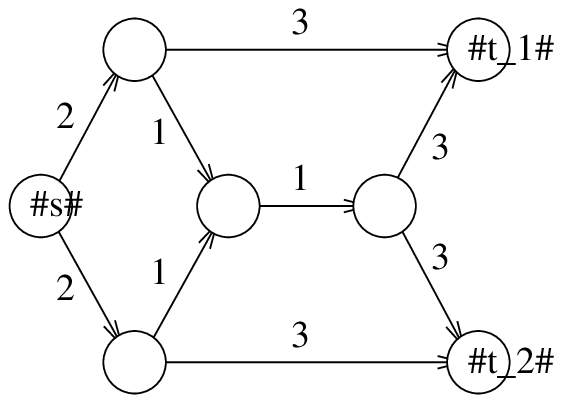} \label{fig:buttcost}}
%\psfrag{#(1/2,1/2,1/2)#}{$(1/2,1/2,1/2)$}
%\psfrag{#(1/2,1/2,0)#}{$(1/2,1/2,0)$}
%\psfrag{#(1/2,0,1/2)#}{$(1/2,0,1/2)$}
\psfrag{#(1/2,1/2,1/2)#}{\small $(\frac{1}{2},\frac{1}{2},\frac{1}{2})$}
\psfrag{#(1/2,1/2,0)#}{\small $(\frac{1}{2},\frac{1}{2},0)$}
\psfrag{#(1/2,0,1/2)#}{\small $(\frac{1}{2},0,\frac{1}{2})$}
\subfigure[Each arc is marked with the triple $(z_{ij}, x_{ij}^{(1)},
x_{ij}^{(2)})$.]
{\includegraphics[scale=1.2]{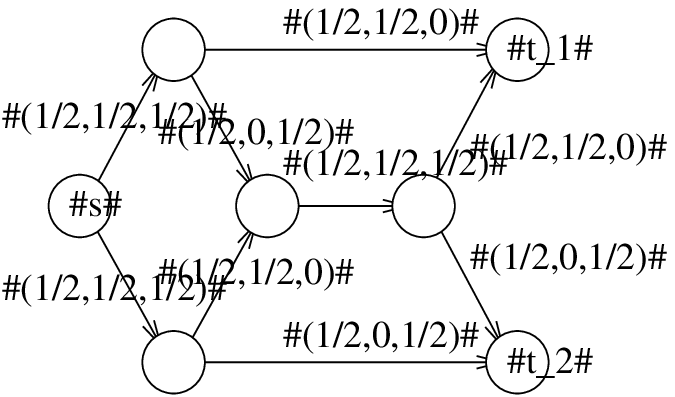} \label{fig:buttcost-1}}
\psfrag{#X_1#}{\small $X_1$}
\psfrag{#X_2#}{\small $X_2$}
\psfrag{#X_1+X_2#}{\small $X_1+X_2$}
\subfigure[Each arc is marked with its code.]
{\includegraphics[scale=1.2]{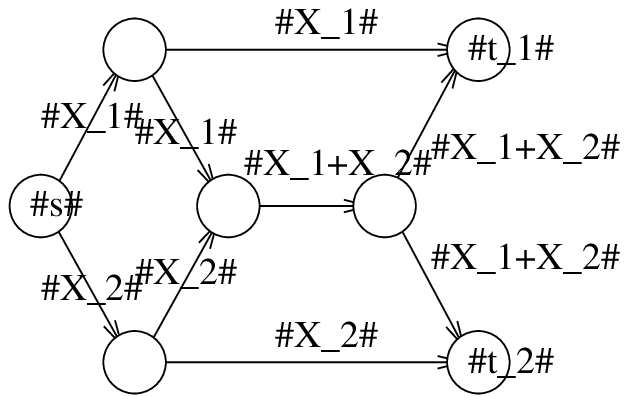} \label{fig:butterfly}}
\caption{A network with multicast from $s$ to $T=\{t_1, t_2\}$.}
\end{figure}

As an example,
consider the network depicted in
Figure~\ref{fig:buttcost}.  We wish to achieve
multicast of unit rate
to two sinks, $t_1$ and $t_2$. 
We have $Z = [0,1]^{|\mathcal{A}|}$ and $f(z) = \sum_{(i,j) \in
\mathcal{A}} a_{ij}z_{ij}$,
where $a_{ij}$ is the cost per unit rate shown beside each link.
An optimal solution to problem (\ref{eqn:1}) for this network is shown
in Figure~\ref{fig:buttcost-1}.
We have flows, $x^{(1)}$ and $x^{(2)}$, of unit size 
from $s$ to $t_1$ and
$t_2$, respectively and, for each arc $(i,j)$, $z_{ij} = 
\max(x_{ij}^{(1)}, x_{ij}^{(2)})$, 
as we expect from the optimization.
To achieve the optimal cost, we code over the subgraph $z$.
A code of length 2 for the subgraph is given in \cite[Figure 7]{acl00},
which we reproduce in
Figure~\ref{fig:butterfly}.  In the figure, $X_1$ and $X_2$ refer to the 
two packets in a coding block.  The coding that is performed is that one of
the interior nodes receives both $X_1$ and $X_2$ and forms the binary
sum of the two, outputting the packet $X_1 + X_2$.  The
code allows both $t_1$ and $t_2$ to recover both $X_1$ and $X_2$ and
it achieves a cost of $19/2$. 

Given a solution of problem (\ref{eqn:1}), there are various coding
schemes that can be used to realize the connection.  The schemes
described in \cite{cwj03, lmk} operate continuously, with each node
continually sending out packets as causal functions of received packets.
The schemes described in
\cite{acl00, lyc03, kom03, jsc05, hmk}, on the other hand, operate in a
block-by-block manner, with each node sending out a block of packets as
a function of its received block.  
In the latter case, the delay
incurred by each arc's block is upper bounded by $\delta/R$ for some
non-negative integer $\delta$ provided that $z_{ij} / R \in
\mathbb{Z}/\delta$ for all $(i,j) \in \mathcal{A}$.  We unfortunately
cannot place such constraints into problem (\ref{eqn:1}) since they
would make it prohibitively difficult.  An alternative is, given $z$, to
take $\lceil \delta z / R \rceil R / \delta$ as the subgraph instead.
Since $\lceil \delta z / R \rceil R / \delta < (\delta z / R +
1)R/\delta = z + R/\delta$, we can guarantee that $\lceil \delta z / R
\rceil R / \delta$ lies in the constraint set $Z$ by looking at $z +
R/\delta$ instead of $z$, resulting in the optimization problem
\begin{equation}
\begin{split}
& \begin{aligned}
\text{minimize }   & f(z + R/\delta) \\
\text{subject to } 
& z + R/\delta \in Z,  \\
& z_{ij} \ge x_{ij}^{(t)} \ge 0,
  \qquad \text{$\forall$ $(i,j) \in A$, $t \in T$} , 
\end{aligned} \\
&\; \sum_{\{j | (i,j) \in A\}} x_{ij}^{(t)}
  - \sum_{\{j | (j,i) \in A\}} x_{ji}^{(t)} 
= \sigma_i^{(t)}, \\
& \qquad \qquad \qquad \qquad \qquad \qquad
\qquad \text{$\forall$ $i \in N$, $t \in T$} . 
\end{split}
\label{eqn:1a}
\end{equation}
We see that, by suitable redefinition of $f$ and $Z$, problem
(\ref{eqn:1a}) can be reduced to problem (\ref{eqn:1}).  Hence, in the
remainder of the paper, we focus only on problem (\ref{eqn:1}).

\subsection{Linear, separable cost and separable constraints}
\label{sec:linear}

The case of linear, separable cost and separable constraints addresses
scenarios where a fixed cost (e.g.,\  monetary cost, energy cost, or
imaginary weight cost) is paid per unit rate placed on an arc and each
arc is subject to a separate constraint (the 
closed interval from 0 to some non-negative capacity).  This is the case
in the network depicted in Figure~\ref{fig:buttcost}.  So,
with each arc $(i,j)$, we associate non-negative numbers $a_{ij}$
and $c_{ij}$, which are the cost per unit rate and the capacity of the
arc, respectively.  Hence, the optimization problem (\ref{eqn:1})
becomes the following linear optimization problem.
\begin{equation}
\begin{split}
& \begin{aligned}
\text{minimize }   & \sum_{(i,j) \in A} a_{ij}z_{ij} \\
\text{subject to } 
& c_{ij} \ge z_{ij}, \qquad \text{$\forall$ $(i,j) \in A$},  \\
& z_{ij} \ge x_{ij}^{(t)} \ge 0,
  \qquad \text{$\forall$ $(i,j) \in A$, $t \in T$} , 
\end{aligned} \\
&\; \sum_{\{j | (i,j) \in A\}} x_{ij}^{(t)}
  - \sum_{\{j | (j,i) \in A\}} x_{ji}^{(t)} 
= \sigma_i^{(t)}, \\
& \qquad \qquad \qquad \qquad \qquad \qquad
\qquad \text{$\forall$ $i \in N$, $t \in T$} . 
\end{split}
\label{eqn:2}
\end{equation}

Unfortunately, the linear optimization problem (\ref{eqn:2}) as it
stands requires centralized computation with full knowledge of the
network.  Motivated by successful network algorithms such as 
distributed Bellman-Ford \cite[Section 5.2]{beg92}, 
we seek a decentralized method 
for solving problem (\ref{eqn:2}), which, when married with
decentralized schemes for constructing network codes \cite{hmk,
lmk, cwj03}, results in a
fully decentralized approach for achieving minimum-cost multicast in the
case of linear, separable cost and separable constraints.  

Toward the end of developing such an algorithm,
we consider the Lagrangian dual problem 
\begin{equation}
\begin{split}
\text{maximize }   & \sum_{t \in T} q^{(t)}(p^{(t)})  \\
\text{subject to } & \sum_{t \in T} p_{ij}^{(t)} = a_{ij} , 
  \qquad \text{$\forall$ $(i,j) \in \mathcal{A}$} , \\
& p_{ij}^{(t)} \ge 0 ,
  \qquad \text{$\forall$ $(i,j) \in \mathcal{A}$, $t \in T$} , 
\end{split}
\label{eqn:12}
\end{equation}
where 
\begin{equation}
q^{(t)}(p^{(t)}) := \min_{x^{(t)} \in F^{(t)}}
  \sum_{(i,j) \in \mathcal{A}} p_{ij}^{(t)} x_{ij}^{(t)} ,
\label{eqn:13}
\end{equation}
and $F^{(t)}$ is the bounded polyhedron of 
points $x^{(t)}$ satisfying 
the conservation of flow constraints
\[
\sum_{\{j | (i,j) \in \mathcal{A}\}} x_{ij}^{(t)} 
- \sum_{\{j | (j,i) \in \mathcal{A}\}} x_{ji}^{(t)} = \sigma_i^{(t)},
\quad \text{$\forall$ $i \in \mathcal{N}$} ,
\]
and capacity constraints 
\[
0 \le x_{ij}^{(t)} \le c_{ij}, 
\quad \text{$\forall$ $(i,j) \in \mathcal{A}$} .
\]

Subproblem (\ref{eqn:13}) is a standard linear minimum-cost 
flow problem,
which can be solved using a multitude of different methods (see, for
example, \cite[Chapters 4--7]{ber98} or \cite[Chapters 9--11]{amo93});
in particular, it can be solved in an asynchronous, distributed manner
using the 
$\varepsilon$-relaxation method \cite[Sections 5.3 and 6.5]{bet89}.
In addition, if the connection rate is small compared to the arc
capacities (more precisely, if $R \le c_{ij}$ for all $(i,j) \in
\mathcal{A}$), 
then subproblem (\ref{eqn:13}) reduces to a shortest path problem, 
which admits a simple, asynchronous, distributed solution \cite[Section
5.2]{beg92}.

Now, to solve the dual problem (\ref{eqn:12}), we employ subgradient
optimization (see, for example, \cite[Section 6.3.1]{ber95} or 
\cite[Section I.2.4]{new99}).  We start
with an iterate $p[0]$ in the feasible set of
(\ref{eqn:12}) and, given an iterate $p[n]$ for some
non-negative integer $n$, we solve
subproblem (\ref{eqn:13}) for each $t$ in $T$ to obtain 
$x[n]$.  We then assign
\begin{equation}
p_{ij}[n + 1] :=
\argmin_{v \in P_{ij}}
\sum_{t \in T} 
(v^{(t)} - (p_{ij}^{(t)}[n] + \theta[n]x_{ij}^{(t)}[n]))^2 
\label{eqn:16}
\end{equation}
for each $(i,j) \in \mathcal{A}$,
where $P_{ij}$ is the $|T|$-dimensional simplex
\[
P_{ij} = 
\left\{v \left| 
\sum_{t \in T} v^{(t)} = a_{ij},\,
v \ge 0 \right.\right\}
\]
and $\theta[n] > 0$ is an appropriate step size.
Thus, $p_{ij}[n + 1]$ is set to be the Euclidean projection of
$p_{ij}[n] + \theta[n]x_{ij}[n]$ onto $P_{ij}$.

To perform the projection, we use the following algorithm, the
justification of which we defer to Appendix~\ref{app:1}.  
Let $u := p_{ij}[n] +
\theta[n]x_{ij}[n]$ and suppose we index the elements of $T$ such
that $u^{(t_1)} \ge u^{(t_2)} \ge \ldots \ge u^{(t_{|T|})}$.
Take $\hat{k}$ to be the smallest $k$ such that
\[
\frac{1}{k}\left(a_{ij} - \sum_{r=1}^{t_k} u^{(r)} \right) \le -u^{(t_{k+1})}
\]
or set $\hat{k} = |T|$ if no such $k$ exists.
Then the projection is achieved by 
\[
p_{ij}^{(t)}[n+1] 
=  
\begin{cases}
u^{(t)} + \frac{a_{ij} - \sum_{r=1}^{t_{\hat{k}}} u^{(r)}}{\hat{k}}
  & \text{if $t \in \{t_{1}, \ldots, t_{\hat{k}}\}$}, \\
0 & \text{otherwise} .
\end{cases}
\]

The disadvantage of subgradient optimization is that, whilst it yields
good approximations of the 
optimal value of the Lagrangian dual problem (\ref{eqn:12}) after
sufficient iteration, it does not necessarily yield a primal optimal
solution.  There are, however, methods for recovering primal solutions
in subgradient optimization.  
We employ the following method, which is due to Sherali and
Choi \cite{shc96}.

Let $\{\mu_l[n]\}_{l = 1,\ldots, n}$ be a sequence of 
convex combination weights for each non-negative integer $n$, 
i.e.\  $\sum_{l=1}^n \mu_l[n] = 1$ and $\mu_l[n] \ge 0$ for all 
$l = 1, \ldots, n$.
Further, let us define
\[
\gamma_{ln} := \frac{\mu_l[n]}{\theta[n]},
\quad\text{$l = 1,\ldots, n$, $n = 0, 1, \ldots$} ,
\]
and 
\[
\Delta \gamma_n^{\max} := \max_{l = 2, \ldots, n}
\{\gamma_{ln} - \gamma_{(l-1)n}\} .
\]
If the step sizes $\{\theta[n]\}$ and convex combination weights
$\{\mu_l[n]\}$ are chosen such that
\begin{enumerate}
\item $\gamma_{ln} \ge \gamma_{(l-1)n}$ for all $l = 2, \ldots, n$ and
$n = 0, 1, \ldots$,
\item $\Delta \gamma_n^{\max} \rightarrow 0$ as $n \rightarrow \infty$,
and
\item $\gamma_{1n} \rightarrow 0$ as $n \rightarrow \infty$ and
$\gamma_{nn} \le \delta$ for all $n = 0,1,\ldots$ for some $\delta > 0$,
\end{enumerate}
then 
we obtain an optimal solution to the primal problem (\ref{eqn:2}) from
any accumulation point of the sequence of primal iterates
$\{\tilde{x}[n]\}$ given by
\begin{equation}
\tilde{x}[n] := \sum_{l=1}^n\mu_l[n]x[l], 
\quad n = 0,1, \ldots .
\label{eqn:17}
\end{equation}
We justify this primal recovery method in Appendix~\ref{app:1}.

The required conditions on the step sizes and convex combination
weights are satisfied by the following choices
\cite[Corollaries 2--4]{shc96}:
\begin{enumerate}
\item step sizes $\{\theta[n]\}$ such that 
$\theta[n] > 0$, $\lim_{n \rightarrow 0} \theta[n] = 0$,
$\sum_{n = 1}^\infty \theta_n = \infty$, and convex combination weights
$\{\mu_l[n]\}$ given by 
$\mu_l[n] = \theta[l] / \sum_{k=1}^n\theta[k]$ for all 
$l = 1, \ldots, n$, $n = 0,1,\ldots$;
\item step sizes $\{\theta[n]\}$ given by 
$\theta[n] = a/(b + cn)$ for all $n = 0,1,\ldots$,
where $a > 0$, $b \ge 0$ and $c > 0$,
and convex combination weights $\{\mu_l[n]\}$ given by
$\mu_l[n] = 1/n$ for all
$l = 1, \ldots, n$, $n = 0,1,\ldots$; and 
\item step sizes $\{\theta[n]\}$ given by 
$\theta[n] = n^{-\alpha}$ for all $n = 0,1,\ldots$, where 
$0 < \alpha < 1$, 
and convex combination weights $\{\mu_l[n]\}$ given by
$\mu_l[n] = 1/n$ for all
$l = 1, \ldots, n$, $n = 0,1,\ldots$.
\end{enumerate}
Moreover, for all three choices, we have $\mu_l[n+1]/\mu_l[n]$
independent of $l$ for all $n$, so primal iterates can be computed
iteratively using
\[
\begin{split}
\tilde{x}[n] &= \sum_{l=1}^n\mu_l[n]x[l] \\
&= \sum_{l=1}^{n-1}\mu_l[n]x[l] + \mu_n[n]x[n] \\
&= \phi[n-1]\tilde{x}[n-1] + \mu_n[n]x[n] ,
\end{split}
\]
where $\phi[n] := \mu_l[n+1]/\mu_l[n]$.

We now have a relatively simple algorithm for 
computing optimal feasible solutions to problem (\ref{eqn:2}) in a
decentralized manner, with computation taking place at each node, which
needs only to be aware of the capacities and costs of its incoming and
outgoing arcs.  
For example, for all arcs $(i,j)$ in $\mathcal{A}$, we can set $p_{ij}^{(t)}[0] =
a_{ij}/|T|$ at both nodes $i$ and $j$.
Since each node has the capacities and costs of its incoming and
outgoing arcs for subproblem (\ref{eqn:13}) for each $t \in T$, we can
apply the $\varepsilon$-relaxation method to obtain
flows $x^{(t)}[0]$ for each $t \in T$, which we use to compute
$p_{ij}[1]$ and $\tilde{x}_{ij}[0]$ at both nodes $i$ and $j$ using
equations (\ref{eqn:16}) and (\ref{eqn:17}), respectively.
We then re-apply the $\varepsilon$-relaxation method and so on.

Although the decentralized algorithm that we have just discussed could
perhaps be extended to convex cost functions (by modifying the dual
problem and employing the $\varepsilon$-relaxation method for convex
cost network flow problems \cite{bpt97, lmz99}), a significantly more
direct and natural method is possible, which we proceed to present.

\subsection{Convex, separable cost and separable constraints}
\label{sec:convex}

Let us now consider the case where,
rather than a cost per unit rate for each arc, we have a
convex, monotonically increasing cost function $f_{ij}$ for
arc $(i,j)$.  Such cost functions arise naturally when the cost is,
e.g.,~latency or congestion.
The optimization problem (\ref{eqn:1})
becomes the following convex optimization problem.
\begin{equation}
\begin{split}
& \begin{aligned}
\text{minimize }   & \sum_{(i,j) \in A} f_{ij}(z_{ij}) \\
\text{subject to } 
& z_{ij} \ge x_{ij}^{(t)} \ge 0,
  \qquad \text{$\forall$ $(i,j) \in A$, $t \in T$} , 
\end{aligned} \\
&\; \sum_{\{j | (i,j) \in A\}} x_{ij}^{(t)}
  - \sum_{\{j | (j,i) \in A\}} x_{ji}^{(t)} 
= \sigma_i^{(t)}, \\
& \qquad \qquad \qquad \qquad \qquad \qquad
\qquad \text{$\forall$ $i \in N$, $t \in T$} . 
\end{split}
\label{eqn:28}
\end{equation}
Note that the capacity constraints have been 
removed, since they can be enforced by making arcs arbitrarily
costly as their flows approach their respective capacities.
We again seek a decentralized method for solving the
subgraph selection problem.

We note that
$z_{ij} = \max_{t \in T} x_{ij}^{(t)}$
at an optimal solution of problem (\ref{eqn:28})
and that $f_{ij}(\max_{t \in T} x_{ij}^{(t)})$ is
a convex function of $x_{ij}$ 
since a monotonically increasing,
convex function of a convex function is convex.  Hence it follows that
problem (\ref{eqn:28}) can be restated as the following convex
optimization problem.
\begin{equation}
\begin{split}
& \begin{aligned}
\text{minimize }   & \sum_{(i,j) \in A} f_{ij}(z_{ij}) \\
\text{subject to } 
& z_{ij} = \max_{t \in T} x_{ij}^{(t)},
\qquad \text{$\forall$ $(i,j) \in A$}, 
\end{aligned} \\
&\; \sum_{\{j | (i,j) \in A\}} x_{ij}^{(t)}
  - \sum_{\{j | (j,i) \in A\}} x_{ji}^{(t)} 
= \sigma_i^{(t)}, \\
& \qquad \qquad \qquad \qquad \qquad \qquad
\qquad \text{$\forall$ $i \in N$, $t \in T$} , \\
&\; x_{ij}^{(t)} \ge 0,
\qquad \text{$\forall$ $(i,j) \in A$, $t \in T$} . 
\end{split}
\label{eqn:29}
\end{equation}

Unfortunately, the max function is not everywhere differentiable, and
this can pose problems for algorithm design.  
We therefore solve the following
modification of problem (\ref{eqn:29}) where the max norm is replaced by
an $l^n$-norm. This replacement was originally proposed in \cite{des04}.
\begin{equation}
\begin{split}
& \begin{aligned}
\text{minimize }   & \sum_{(i,j) \in A} f_{ij}(z_{ij}^\prime) \\
\text{subject to } 
& z_{ij}^\prime = \left(\sum_{t \in T} (x_{ij}^{(t)})^n
\right)^{1/n},
\qquad \text{$\forall$ $(i,j) \in A$}, 
\end{aligned} \\
&\; \sum_{\{j | (i,j) \in A\}} x_{ij}^{(t)}
  - \sum_{\{j | (j,i) \in A\}} x_{ji}^{(t)} 
= \sigma_i^{(t)}, \\
& \qquad \qquad \qquad \qquad \qquad \qquad
\qquad \text{$\forall$ $i \in N$, $t \in T$} , \\
&\; x_{ij}^{(t)} \ge 0,
\qquad \text{$\forall$ $(i,j) \in A$, $t \in T$} . 
\end{split}
\label{eqn:30}
\end{equation}
We have that $z_{ij}^\prime \ge z_{ij}$ for all $n > 0$
and that $z_{ij}^\prime$ approaches $z_{ij}$ as $n$ approaches infinity.
Thus, we shall assume that $n$ is large and attempt to develop a
decentralized algorithm to solve problem (\ref{eqn:30}).  Note that, 
since $z_{ij}^\prime \ge z_{ij}$, a code with rate $z_{ij}^\prime$ on 
each arc $(i,j)$ exists for any feasible solution.

Problem (\ref{eqn:30}) is a convex multicommodity flow problem.  There
are many algorithms for convex multicommodity flow problems (see
\cite{omv00} for a survey), some of which (e.g.~the algorithms in
\cite{ber80,bgg84}) are well-suited for decentralized implementation.
These algorithms can certainly be used, but, in this paper, we propose
solving problem (\ref{eqn:30}) using a primal-dual algorithm derived
from the primal-dual approach to internet congestion control (see
\cite[Section 3.4]{sri04}).

We restrict ourselves to the case where $\{f_{ij}\}$ are strictly convex.
Since the variable $z_{ij}^\prime$ is a
strictly convex function of $x_{ij}$, it follows
that the objective function for problem (\ref{eqn:30}) is strictly
convex, so the problem admits a unique solution for any integer $n >
0$.  Let $U(x) := -\sum_{(i,j) \in \mathcal{A}} f_{ij}(
(\sum_{t \in T}(x_{ij}^{(t)})^n)^{1/n})$, and
let $(y)_x^+$ for $x \geq 0$ denote the following function of $y$:
\begin{equation*}
(y)_x^+ = \begin{cases}  
    y &  \text{if $x>0$},\\
    \max\{y,0\} &   \text{if $x \le 0$}.
\end{cases}
\end{equation*}
Consider the
following continuous-time primal-dual algorithm: 
\begin{gather}
\dot{x}_{ij}^{(t)}
=
k_{ij}^{(t)}(x_{ij}^{(t)})\left(\frac{\partial U({x})}
{\partial x_{ij}^{(t)}} -
q_{ij}^{(t)}+\lambda_{ij}^{(t)}\right) , \label{eqn:alg1} \\
\dot{p}_i^{(t)} 
=
h_i^{(t)}(p_i^{(t)}) (y_i^{(t)} - \sigma_i^{(t)}) , \\
\dot{\lambda}_{ij}^{(t)} = m_{ij}^{(t)}(\lambda_{ij}^{(t)})
\left(-x_{ij}^{(t)}\right)_{\lambda_{ij}^{(t)}}^+ ,
\label{eqn:alg3}
\end{gather}
where
\begin{gather*}
q_{ij}^{(t)} := p_i^{(t)} - p_j^{(t)} , \\
y_i^{(t)} := 
\sum_{\{j | (i,j) \in \mathcal{A}\}} {x}_{ij}^{(t)}
  - \sum_{\{j | (j,i) \in \mathcal{A}\}} {x}_{ji}^{(t)} ,
\end{gather*}
and
$k_{ij}^{(t)}(x_{ij}^{(t)})>0 $, $h_i^{(t)}(p_i^{(t)}) >0 $, 
and $m_{ij}^{(t)}(\lambda_{ij}^{(t)}) > 0$ are non-decreasing
continuous functions of  $x_{ij}^{(t)}$, $p_i^{(t)}$, 
and $\lambda_{ij}^{(t)}$ respectively. 

\begin{Prop}
The algorithm specified by Equations (\ref{eqn:alg1})--(\ref{eqn:alg3}) 
is globally, asymptotically stable.
\label{prop:stability}
\end{Prop}

\begin{proof}
See Appendix~\ref{app:stability}.
\end{proof}

The global, asymptotic stability of the algorithm implies that no matter
what the initial choice of $({x},{p})$ is, the primal-dual
algorithm will converge to the unique solution of problem
(\ref{eqn:30}).  We have to choose ${\lambda}$, however, with
non-negative entries as the initial choice.

We associate a processor with each arc $(i,j)$ and node $i$.  In a
typical setting where there is one processor at every node, we could
assign the processor at a node to be its own processor as well as the
processor for all its outgoing arcs.  

We assume that the processor for node $i$ keeps 
track of the 
variables $\{p_i^{(t)}\}_{t \in T}$, while the processor for
arc $(i,j)$ keeps track of the 
variables $\{\lambda_{ij}^{(t)}\}_{t \in T}$ and 
$\{x_{ij}^{(t)}\}_{t \in T}$. 
With this assumption,
the algorithm is decentralized in
the following sense:
\begin{itemize}
\item a node processor needs only to exchange information with the
processors for arcs coming in or out of the node; and 
\item an arc processor needs only to exchange information with the
processors for nodes that it is connected to.
\end{itemize}
This fact is evident
from equations (\ref{eqn:alg1})--(\ref{eqn:alg3})
by noting that 
\[
\frac{\partial U({x})}{\partial x_{ij}^{(t)}} =
-f_{ij}(z_{ij}^\prime)
\left(x_{ij}^{(t)}/z_{ij}^\prime\right)^{n-1} .
\]

In implementing the primal-dual algorithm, we must bear the
following points in mind.
\begin{itemize}
\item The primal-dual algorithm in (\ref{eqn:alg1})--(\ref{eqn:alg3})
is a continuous time algorithm.
To discretize the algorithm, we consider time steps $m = 1,2,\ldots$
and replace the derivatives by differences:
\begin{gather*}
\begin{aligned}
&x_{ij}^{(t)}[m+1] 
= {x}_{ij}^{(t)}[m] \\
  &\qquad+ \alpha_{ij}^{(t)}[m]\left(\frac{\partial U(x[m])}
{\partial x_{ij}^{(t)}[m]} -
q_{ij}^{(t)}[m]+\lambda_{ij}^{(t)}[m]\right)
\end{aligned} , \\
p_i^{(t)}[m+1]
= p_i^{(t)}[m] + \beta_i^{(t)}[m] (y_i^{(t)}[m] - \sigma_i^{(t)}) , \\
\lambda_{ij}^{(t)}[m+1] = \lambda_{ij}^{(t)}[m] + \gamma_{ij}^{(t)}[m] \left(-x_{ij}^{(t)}[m]\right)_{\lambda_{ij}^{(t)}[m]}^+ ,
\end{gather*}
where
\begin{gather*}
q_{ij}^{(t)}[m] := p_i^{(t)}[m] - p_j^{(t)}[m] , \\
y_i^{(t)}[m] := \sum_{\{j | (i,j) \in \mathcal{A}\}} x_{ij}^{(t)}[m]
- \sum_{\{j | (j,i) \in \mathcal{A}\}} x_{ji}^{(t)}[m]
,
\end{gather*}
and $\alpha_{ij}^{(t)}[m] >0$, $ \beta_i^{(t)}[m] > 0$, and $\gamma_{ij}^{(t)}[m]
> 0$
can be thought of as step sizes.

\item While the algorithm is guaranteed to converge to the optimum solution, the
    value of the variables at any time instant $m$ is not necessarily a feasible
    solution. A start-up time is required before a feasible solution
    is computed.
\item Unfortunately, the above algorithm is a synchronous algorithm where the
    various processors need to exchange information at regular intervals. It is
    an interesting problem to investigate an asynchronous implementation of the
    primal-dual algorithm.  
\end{itemize}

\subsection{Elastic rate demand}

We have thus far focused on the case of an inelastic rate demand, which
is presumably provided by a separate flow control algorithm.  But this
flow control does not necessarily need to be done separately.  Thus, we
now suppose that the rate demand is elastic and that it is represented
by a utility function that has the same units as the cost function, and
we seek to maximize utility minus cost.  We continue to assume strictly
convex, separable cost and separable constraints.

We associate with the source a utility function $U_r$ such that
$U_r(R)$ is the utility derived by the source when $R$ is the data rate.
The function $U_r$ is assumed to be a strictly concave and increasing.
Hence, in this setup, the problem we address is as follows:
\begin{equation}
\begin{split}
& \begin{aligned}
\text{maximize }   & U(x,R)\\
\text{subject to }
\end{aligned} \\
&\; \sum_{\{j | (i,j) \in A\}} x_{ij}^{(t)}
  - \sum_{\{j | (j,i) \in A\}} x_{ji}^{(t)} 
= \sigma_i^{(t)}, \\
& \qquad \qquad \qquad \qquad \qquad
\qquad \text{$\forall$ $i \in N \setminus \{t\}$, $t \in T$} , \\
&\; R \ge 0, \\
&\; x_{ij}^{(t)} \ge 0,
  \qquad \text{$\forall$ $(i,j) \in A$, $t \in T$} , 
\end{split}
\label{eqn:200}
\end{equation}
where
$U(x,R) := U_r(R) - \sum_{(i,j) \in \mathcal{A}} f_{ij}
(\sum_{t \in T}(x_{ij}^{(t)})^n)^{1/n})$.  In problem (\ref{eqn:200}),
some of the flow constraints have been dropped by making the
observation that the equality constraints at a sink $t$, namely
\begin{equation*}
\sum_{\{j | (t,j) \in \mathcal{A}\}} x_{tj}^{(t)}
  - \sum_{\{j | (j,t) \in \mathcal{A}\}} x_{jt}^{(t)} 
= \sigma_t^{(t)} = -R ,
\end{equation*}
follow from the constraints at the source and at the other nodes.  The
dropping of these constraints is crucial to the proof that the algorithm
presented in the sequel is decentralized.  

This problem can be solved by the following primal-dual algorithm.
\begin{gather*}
\allowdisplaybreaks
\dot{x}_{ij}^{(t)} =
k_{ij}^{(t)}(x_{ij}^{(t)})\left(\frac{\partial U({x, R})}
{\partial x_{ij}^{(t)}} -
q_{ij}^{(t)}+\lambda_{ij}^{(t)}\right) , \\
\dot{R} =
k_R(R)\left(\frac{\partial U(x, R)}{\partial R}
- q_R + \lambda_R \right) , \\
\dot{p}_i^{(t)} =
h_i^{(t)}(p_i^{(t)}) y_i^{(t)} , \\
\dot{\lambda}_{ij}^{(t)} = m_{ij}^{(t)}(\lambda_{ij}^{(t)})
\left(-x_{ij}^{(t)}\right)_{\lambda_{ij}^{(t)}}^+ , \\
\dot{\lambda}_R = m_R(\lambda_R)
\left(-R\right)_{\lambda_R}^+ ,
\end{gather*}
where
\begin{gather*}
q_{ij}^{(t)} := p_i^{(t)} - p_j^{(t)} , \\
q_R := - \sum_{t \in T} p_s^{(t)} , \\
y_i^{(t)} := 
\sum_{\{j | (i,j) \in \mathcal{A}\}} \hat{x}_{ij}^{(t)}
  - \sum_{\{j | (j,i) \in \mathcal{A}\}} \hat{x}_{ji}^{(t)} - \sigma_i^{(t)}.
\end{gather*}
It can be shown using similar arguments as those for
Proposition~\ref{prop:stability} that this algorithm is globally,
asymptotically stable.

In addition,
by letting the source $s$ keep track of the rate $R$, it can be seen
that the algorithm is decentralized.

\section{Wireless packet networks}
\label{sec:wireless}

To model wireless packet networks, we take the model for wireline packet
networks and include the effect of two new factors: link lossiness and
link broadcast.  Link lossiness refers to the dropping or loss of
packets as they are transmitted over a link; and link broadcast refers
to how links, rather than necessarily being point-to-point, may
originate from a single node and reach more than one other node.  Our
model includes networks consisting of lossy point-to-point links and
networks consisting of lossless broadcast links as special cases.

We represent the network with a directed hypergraph $\mathcal{H} =
(\mathcal{N}, \mathcal{A})$, where $\mathcal{N}$ is the set of nodes and
$\mathcal{A}$ is the set of hyperarcs. A hypergraph is a generalization
of a graph, where, rather than arcs, we have hyperarcs.  A hyperarc is a
pair $(i,J)$, where $i$, the start node, is an element of $\mathcal{N}$
and $J$, the set of end nodes, is a non-empty subset of $\mathcal{N}$.
Each hyperarc $(i, J)$ represents a lossy broadcast link from node $i$
to nodes in the non-empty set $J$.  We denote by $z_{iJ}$ the rate at
which coded packets are injected into hyperarc $(i,J)$, and we denote by
$z_{iJK}$ the rate at which packets, injected into hyperarc $(i,J)$, are
received by exactly the set of nodes $K \subset J$.  Hence $z_{iJ} :=
\sum_{K \subset J} z_{iJK}$.  Let
\[
b_{iJK} :=
\frac{\sum_{\{L \subset J | L \cap K \neq \emptyset\}} z_{iJL}}{z_{iJ}}.
\]
The rate vector $z$, consisting of
$z_{iJ}$, $(i,J) \in \mathcal{A}$, is called a subgraph, and we assume
that it must lie within a constraint set $Z$
for, if not, the packet queues associated with one or more hyperarcs
becomes unstable (for examples of constraint sets $Z$ that pertain
specifically to multi-hop wireless networks, see
\cite{crs03,jpp03,jxb03,xjb04,kon05,wcz05}).  We reasonably assume that
$Z$ is a convex subset of the positive orthant containing the origin.
We associate with the network a cost function $f$ (reflecting, for
example, the average latency or energy consumption) that maps valid rate
vectors to real numbers and that we seek to minimize.  

Suppose we have a source node $s$ wishing to transmit packets at a
positive, real rate $R$ to a non-empty set of sink nodes $T$.  Consider
the following optimization problem:
\begin{equation}
\begin{split}
&\begin{aligned}
\text{minimize }   & f(z) \\
\text{subject to } 
& z \in Z, 
\end{aligned} \\
&\; z_{iJ} b_{iJK}
\ge \sum_{j \in K} x_{iJj}^{(t)},
  \qquad \text{$\forall$ $(i,J) \in \mathcal{A}$, $K \subset J$, $t \in
T$}, \\
&\; \sum_{\{J | (i,J) \in \mathcal{A}\}} \sum_{j \in J} x_{iJj}^{(t)}
  - \sum_{\{j | (j,I) \in \mathcal{A}, i \in I\}} x_{jIi}^{(t)} 
= \sigma_i^{(t)}, \\
& \qquad \qquad \qquad \qquad \qquad \qquad
\qquad \text{$\forall$ $i \in \mathcal{N}$, $t \in T$} , \\
&\; x_{iJj}^{(t)} \ge 0, 
\qquad \text{$\forall$ $(i,J) \in \mathcal{A}$, $j \in J$, $t \in T$} .
\end{split}
\label{eqn:400}
\end{equation}

\begin{Thm} 
The vector $z$ is part of a feasible solution for the optimization
problem (\ref{eqn:400}) if and only if there exists a network code that
sets up a multicast connection in the wireless network represented by
hypergraph $\mathcal{H}$ at rate arbitrarily close to $R$ from source
$s$ to sinks in the set $T$ and that injects packets at rate arbitrarily
close to $z_{iJ}$ on each hyperarc $(i,J)$.  
\label{thm:5} 
\end{Thm}

\begin{proof}
The proof is much the same as that for Theorem~\ref{thm:4}.  But,
instead of Theorem~1 of \cite{acl00}, we use Theorem~2 of
\cite{lmk}.
\end{proof}

In the lossless case, we have $b_{iJK} = 1$ for all non-empty $K \subset
J$ and $b_{iJ\emptyset} = 0$.  Hence, problem (\ref{eqn:400}) simplifies
to the following optimization problem.
\begin{equation}
\begin{split}
&\begin{aligned}
\text{minimize }   & f(z) \\
\text{subject to } 
& z \in Z, 
\end{aligned} \\
&\; z_{iJ} 
\ge \sum_{j \in J} x_{iJj}^{(t)},
  \qquad \text{$\forall$ $(i,J) \in \mathcal{A}$, $t \in T$}, \\
&\; \sum_{\{J | (i,J) \in \mathcal{A}\}} \sum_{j \in J} x_{iJj}^{(t)}
  - \sum_{\{j | (j,I) \in \mathcal{A}, i \in I\}} x_{jIi}^{(t)} 
= \sigma_i^{(t)}, \\
& \qquad \qquad \qquad \qquad \qquad \qquad
\qquad \text{$\forall$ $i \in \mathcal{N}$, $t \in T$} , \\
&\; x_{iJj}^{(t)} \ge 0, 
\qquad \text{$\forall$ $(i,J) \in \mathcal{A}$, $j \in J$, $t \in T$} .
\end{split}
\label{eqn:422}
\end{equation}

A simplification of problem (\ref{eqn:422}) can be made if we assume
that, when nodes transmit in a lossless network, they reach all nodes in
a certain area, with cost increasing as this area is increased.  More
precisely, suppose that we have separable cost, so $f(z) = \sum_{(i,J)
\in \mathcal{A}} f_{iJ}(z_{iJ})$.  Suppose further that
each node $i$ has $M_i$ outgoing hyperarcs
$(i,J_1^{(i)}), (i, J_2^{(i)}), \ldots, (i, J_{M_i}^{(i)})$
with 
$J_1^{(i)} \subsetneq J_2^{(i)} \subsetneq \cdots \subsetneq
J_{M_i}^{(i)}$.
(We assume that there are no identical links, as duplicate links
can effectively be treated as a single link.)
Then, we assume that 
 $f_{iJ_1^{(i)}}(\zeta) < f_{iJ_2^{(i)}}(\zeta) < \cdots
< f_{iJ_{M_i}^{(i)}}(\zeta)$ for all $\zeta \ge 0$ and
nodes $i$.  
For $(i,j) \in \mathcal{A}^\prime
:= \{(i,j) | (i,J) \in A, J \ni j\}$,
we introduce the variables
\[
\hat{x}_{ij}^{(t)} := \sum_{m=m(i, j)}^{M_i} x_{iJ_m^{(i)}j}^{(t)},
\]
where $m(i,j)$ is the unique $m$ such that $j \in J_m^{(i)} \setminus
J_{m-1}^{(i)}$
(we define $J_0^{(i)} := \emptyset$ for all $i \in \mathcal{N}$
for convenience).  Now,
problem (\ref{eqn:422}) can be reformulated as the following
optimization problem, which has substantially fewer variables.
\begin{equation}
\begin{split}
&\begin{aligned}
\text{minimize }   & \sum_{(i,J) \in \mathcal{A}} f_{iJ}(z_{iJ}) \\
\text{subject to } 
& z \in Z, 
\end{aligned} \\
&\; \sum_{n=m}^{M_i} z_{iJ_n^{(i)}} 
\ge \sum_{k \in J_{M_i}^{(i)} \setminus J_{m-1}^{(i)}}
\hat{x}_{ik}^{(t)}, \\
& \qquad \qquad \qquad 
  \qquad \text{$\forall$ $i \in \mathcal{N}$, $m = 1, \ldots, M_i$, $t \in T$}, \\
&\; \sum_{\{j | (i,j) \in \mathcal{A}^\prime\}} \hat{x}_{ij}^{(t)}
  - \sum_{\{j | (j,i) \in \mathcal{A}^\prime\}} \hat{x}_{ji}^{(t)} 
= \sigma_i^{(t)}, \\
& \qquad \qquad \qquad \qquad \qquad \qquad
\qquad \text{$\forall$ $i \in \mathcal{N}$, $t \in T$} , \\
&\; \hat{x}_{ij}^{(t)} \ge 0, 
\qquad \text{$\forall$ $(i,j) \in \mathcal{A}^\prime$, $t \in T$} .
\end{split}
\label{eqn:450}
\end{equation}

\begin{Prop}
Suppose that $f(z) = \sum_{(i,J) \in \mathcal{A}}f_{iJ}(z_{ij})$
and that $f_{iJ_1^{(i)}}(\zeta) < f_{iJ_2^{(i)}}(\zeta) < \cdots
< f_{iJ_{M_i}^{(i)}}(\zeta)$ for all $\zeta \ge 0$ and
nodes $i$.
Then
problem (\ref{eqn:422}) and problem (\ref{eqn:450}) are equivalent in
the sense that they have the same optimal cost and $z$ is part of an
optimal solution for (\ref{eqn:422}) if and only if it is part of an
optimal solution for (\ref{eqn:450}).
\label{prop:10}
\end{Prop}

\begin{proof}
See Appendix~\ref{app:prop_10}.
\end{proof}

We see that, provided that $\{b_{iJK}\}$ are constant, problems
(\ref{eqn:400}) and (\ref{eqn:422}) are of essentially the same form as
problem (\ref{eqn:1}), albeit with possibly more linear constraints
relating $z$ and $x$, and, if we drop the constraint set $Z$ and
consider linear, separable cost or convex, separable cost, then the
decentralized algorithms discussed in Sections~\ref{sec:linear}
and~\ref{sec:convex} can be applied with little modification.
In the case of problem (\ref{eqn:450}), the subgradient method of
Section~\ref{sec:linear} can be applied once we note that its Lagrangian
dual,
\begin{equation*}
\begin{split}
\text{maximize }   & \sum_{t \in T} \hat{q}^{(t)}(p^{(t)})  \\
\text{subject to } & \sum_{t \in T}
p_{iJ_m^{(i)}}^{(t)} = s_{iJ_m^{(i)}} ,
  \qquad \text{$\forall$ $i \in \mathcal{N}$, $m=1,\ldots,M_i$} , \\
& p_{iJ}^{(t)} \ge 0 ,
  \qquad \text{$\forall$ $(i,J) \in \mathcal{A}$, $t \in T$} ,
\end{split}
\end{equation*}
where
\[
s_{iJ_m^{(i)}} := a_{iJ_m^{(i)}} - a_{iJ_{m-1}^{(i)}},
\]
and
\begin{equation*}
\hat{q}^{(t)}(p^{(t)}) := \min_{\hat{x}^{(t)} \in \hat{F}^{(t)}}
  \sum_{(i,j) \in \mathcal{A}^\prime} \left(\sum_{m=1}^{m(i,j)}
p_{iJ_m^{(i)}}^{(t)}\right) \hat{x}_{ij}^{(t)} ,
\end{equation*}
is of the same form as (\ref{eqn:12}).

\section{Comparison with techniques in routed packet networks}
\label{sec:comparison}

In this section, we report on the results of several simulations that we
conducted to assess the performance of the proposed techniques.  We
begin with wireline networks.

\begin{table*}
\centering
\begin{tabular}{|c|c|c|c|c|c|} \hline
Network & Approach &
\multicolumn{4}{c|}{Average multicast cost} \\ \cline{3-6}
& & 2 sinks & 4 sinks & 8 sinks & 16 sinks \\ \hline
Telstra (au) & DST approximation &
17.0 & 28.9 & 41.7 & 62.8 \\
& Network coding &
13.5 & 21.5 & 32.8 & 48.0 \\ \hline
Sprint (us) & DST approximation &
30.2 & 46.5 & 71.6 & 127.4 \\ 
& Network coding &
22.3 & 35.5 & 56.4 & 103.6 \\ \hline
Ebone (eu) & DST approximation & 
28.2 & 43.0 & 69.7 & 115.3 \\
& Network coding &
20.7 & 32.4 & 50.4 & 77.8 \\ \hline
Tiscali (eu) & DST approximation &
32.6 & 49.9 & 78.4 & 121.7 \\
& Network coding &
24.5 & 37.7 & 57.7 & 81.7 \\ \hline
Exodus (us) & DST approximation &
43.8 & 62.7 & 91.2 & 116.0 \\
& Network coding &
33.4 & 49.1 &  68.0 &  92.9 \\ \hline 
Abovenet (us) & DST approximation &
27.2 & 42.8 & 67.3 & 75.0 \\
& Network coding &
21.8 & 33.8 & 60.0 & 67.3 \\ \hline
\end{tabular}
\caption{Average cost of random multicast connections of unit rate
for various approaches
in graphs representing various ISP networks.  The cost per unit rate 
on each arc is the link weight as assessed by the Rocketfuel project of
the University of Washington \cite{msw02}.  
Source and sink nodes were
selected according to a uniform distribution over all possible
selections.}
\label{tab:1}
\end{table*}

In routed wireline networks, the standard approach to establishing
minimum-cost multicast connections is to find the shortest tree rooted
at the source that reaches all the sinks, which equates to solving the
Steiner tree problem on directed graphs \cite{ram96}.  For coded
networks, the analogous problem to finding the shortest tree is solving
the linear optimization problem (\ref{eqn:2}) in the case where $c_{ij}
= +\infty$, which, being a linear optimization problem, admits a
polynomial-time solution.  By contrast, the Steiner tree problem on
directed graphs is well-known to be NP-complete.  Although tractable
approximation algorithms exist for the Steiner tree problem on directed
graphs (for example, \cite{ram96, ccc99, zok02}), the solutions thus
obtained are suboptimal relative to minimum-cost multicast without
coding, which in turn is suboptimal relative to when coding is used,
since coding subsumes forwarding and replicating (for example, the
optimal cost for a Steiner tree in the network in
Figure~\ref{fig:buttcost} is 10, as opposed to $19/2$).  Thus, coding
promises potentially significant cost improvements.  

We conducted simulations where we took graphs representing various
Internet Service Provider (ISP) networks and assessed the average total
weight of random multicast connections using, first, our proposed
network-coding based solution and, second, routing over the tree given
by the Directed Steiner Tree (DST) approximation algorithm described
in~\cite{ccc99}.  The graphs, and their associated link weights, were
obtained from the Rocketfuel project of the University of Washington
\cite{msw02}.  The approximation algorithm in \cite{ccc99} was chosen
for comparison as it achieves a poly-logarithmic approximation ratio (it
achieves an approximation ratio of $O(\log^2|T|)$, where $|T|$ is the
number of sink nodes), which is roughly as good as can be expected from
any practical algorithm, since it has been shown that it is highly
unlikely that there exists a polynomial-time algorithm that can achieve
an approximation factor smaller than logarithmic \cite{ram96}.  The
results of the simulations are tabulated in Table~\ref{tab:1}.  We see
that, depending on the network and the size of the multicast group, the
average cost reduction ranges from 10\% to 33\%.  Though these
reductions are modest, it is important to keep in mind that our proposed
solution easily accommodates decentralized operation.

\begin{table*}
\centering
\begin{tabular}{|c|c|c|c|c|c|} \hline
Network size & Approach &
\multicolumn{4}{c|}{Average multicast energy} \\ \cline{3-6}
& & 2 sinks & 4 sinks & 8 sinks & 16 sinks \\ \hline
20 nodes & MIP algorithm &
30.6 & 33.8 & 41.6 & 47.4 \\
& Network coding &
15.5 &  23.3 &  29.9 &  38.1 \\ \hline
30 nodes & MIP algorithm &
26.8 & 31.9 & 37.7 & 43.3 \\
& Network coding &
15.4 &  21.7 &  28.3 &  37.8 \\ \hline
40 nodes & MIP algorithm &
24.4 & 29.3 & 35.1 & 42.3 \\
& Network coding &
14.5 &  20.6 &  25.6 &  30.5 \\ \hline
50 nodes & MIP algorithm &
22.6 & 27.3 & 32.8 & 37.3 \\
& Network coding &
12.8 &  17.7 &  25.3 &  30.3 \\ \hline
\end{tabular}
\caption{Average energy of random multicast connections of unit rate
for various approaches
in random wireless networks of varying size.
Nodes were placed randomly within a $10 \times 10$ square with
a radius of connectivity of 3.
The energy required to transmit at rate $z$
to a distance $d$ was taken to be $d^2z$.
Source and sink nodes were
selected according to a uniform distribution over all possible
selections.}
\label{tab:2}
\end{table*}

For wireless networks,
one specific problem of interest is that of minimum-energy multicast
(see, for example, \cite{wne02b,lia02}).  In this problem, we wish to
achieve minimum-energy multicast in a lossless wireless network without
explicit regard for throughput or bandwidth, so the constraint set $Z$
can be dropped altogether.
The cost function is linear and separable, namely, it is
$f(z) = \sum_{(i,J) \in \mathcal{A}} a_{iJ}z_{iJ}$, where
$a_{iJ}$ represents the energy required to transmit a packet
to nodes in $J$ from node $i$.
Hence problem
(\ref{eqn:450}) becomes a linear optimization problem with a polynomial
number of constraints, which can therefore be solved in polynomial time.
By contrast, the same problem using traditional routing-based
approaches is NP-complete---in fact, the special case of broadcast in
itself is NP-complete, a result shown in \cite{lia02, ams02}.
The problem must therefore be
addressed using polynomial-time heuristics such as the
MIP algorithm proposed in \cite{wne02b}.

We conducted simulations where we placed nodes randomly, according to a
uniform distribution, in a $10 \times
10$ square with a radius of connectivity of 3 and assessed the average
total energy of random multicast connections using first, our proposed
network-coding based solution and, second, the routing solution given by
the MIP algorithm.  
The energy required to transmit at rate $z$
to a distance $d$ was taken to be $d^2z$.
The results of the simulations are tabulated in Table~\ref{tab:2}.  We
see that, depending on the size of the network and the size of the
multicast group, the average energy reduction ranges from 13\% to 49\%.
These reductions are more substantial than those for the wireline
simulations, but are still modest.  Again, it is important to keep in
mind that the proposed solution easily accommodates decentralized
operation.

\begin{figure}
\begin{center}
\includegraphics[height=2in]{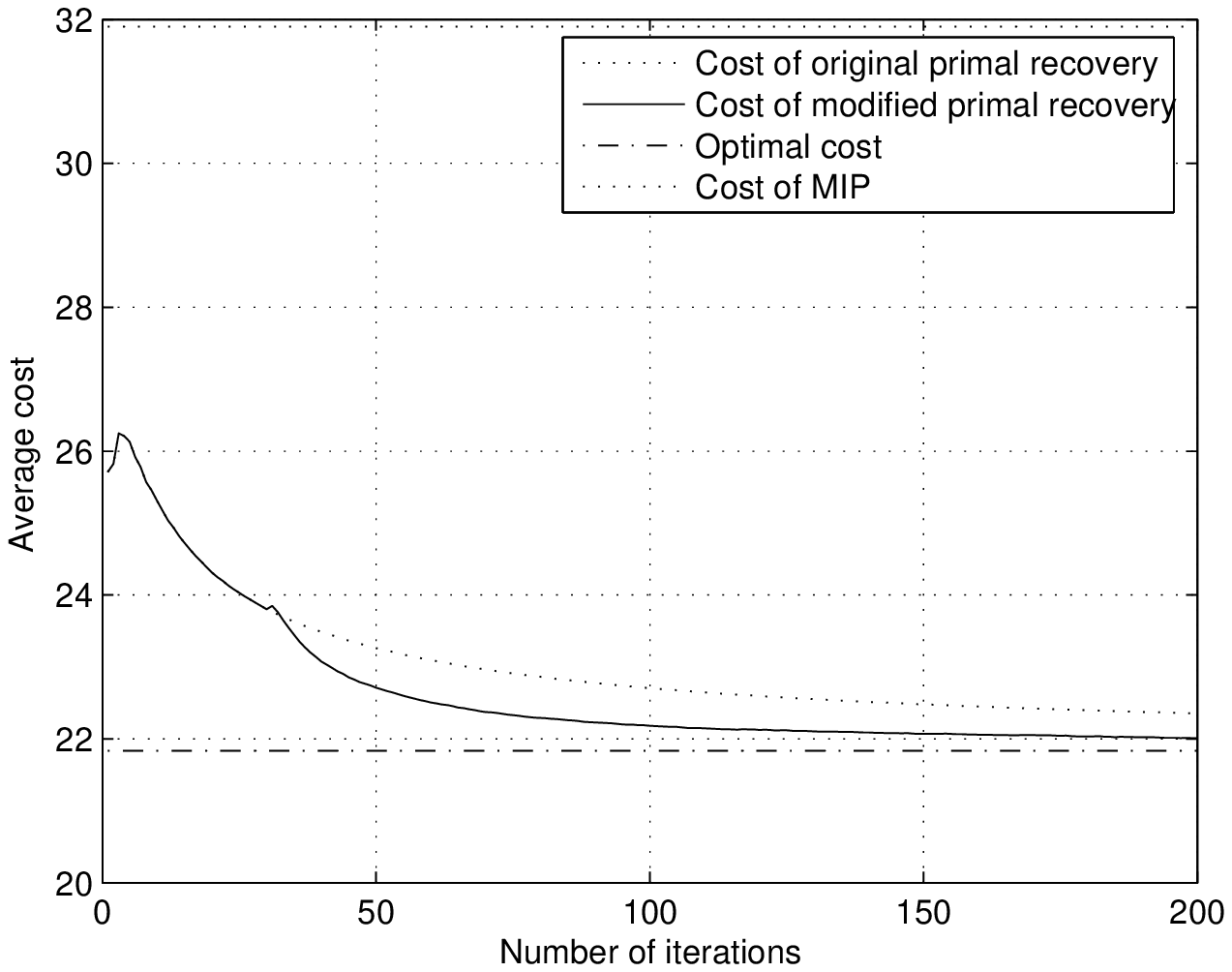}
\end{center}
\caption{Average energy of a random 4-terminal multicast of unit rate 
in a 30-node wireless network using the subgradient method of
Section~\ref{sec:linear}.
Nodes were placed randomly within a $10 \times 10$ square with
a radius of connectivity of 3.
The energy required to transmit at rate $z$
to a distance $d$ was taken to be $d^2z$.
Source and sink nodes were
selected according to a uniform distribution over all possible
selections.}
\label{fig:fig1a}
\end{figure}

\begin{figure}
\begin{center}
\includegraphics[height=2in]{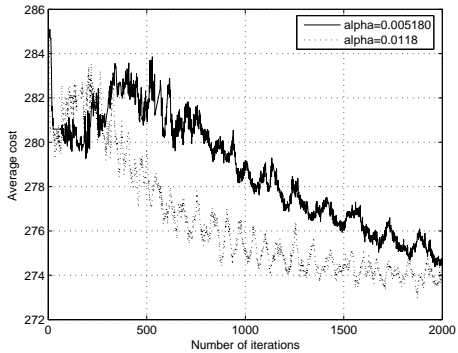}
\end{center}
\caption{Average energy of a random 4-terminal multicast of unit rate 
in a 30-node wireless network using the primal-dual method of
Section~\ref{sec:convex}.
Nodes were placed randomly within a $10 \times 10$ square with
a radius of connectivity of 3.
The energy required to transmit at rate $z$
to a distance $d$ was taken to be $d^2 e^z$.
Source and sink nodes were
selected according to a uniform distribution over all possible
selections.}
\label{fig:nayakfig}
\end{figure}

We conducted simulations on our decentralized algorithms 
for a network of 30 nodes and a multicast group of 4 terminals
under the same set up.
In Figure~\ref{fig:fig1a}, we show the average behavior of the
subgradient method of Section~\ref{sec:linear} applied to problem
(\ref{eqn:450}).  The algorithm was run under two choices of step sizes
and convex combination weights.  The curve labeled ``original primal
recovery'' refers to the case where the step sizes are given by
$\theta[n] = n^{-0.8}$ and the convex combination weights by $\mu_l[n] =
1/n$.  The curve labeled ``modified primal recovery'' refers to the case
where the step sizes are given by $\theta[n] = n^{-0.8}$ and the convex
combination weights by $\mu_l[n] = 1/n$, if $n < 30$, and $\mu_l[n] =
1/30$, if $n \ge 30$.  The modified primal recovery rule was chosen as a
heuristic to lessen the effect of poor primal solutions obtained in
early iterations.  For reference, the optimal cost of problem
(\ref{eqn:450}) is shown, as is the cost obtained by the MIP algorithm.
We see that, for both choices of step sizes and convex combination
weights, the cost after the first iteration is already lower than that
from the MIP algorithm.  Moreover, in fewer than 50 iterations, the cost
using modified primal recovery is within $5\%$ of the
optimal value.  Thus, in a small number of iterations, the
subgradient method yields significantly lower energy consumption than
that obtained by the MIP algorithm, which is centralized.

In Figure~\ref{fig:nayakfig}, we show the average behavior of the
primal-dual method of Section~\ref{sec:convex} applied to problem
(\ref{eqn:422}).  To make the cost strictly convex, the energy required
to transmit at rate $z$ to a distance $d$ was taken to be $d^2e^z$.
Recall that we do not necessarily have a feasible solution at each
iteration.  Thus, to compare the cost at the end of each iteration, we
recover a feasible solution from the vector $z^\prime[m]$ as follows: We
take the subgraph defined by $z^\prime[m]$ and compute the maximum flow
from source $s$ to sinks in the set $\mathcal{T}$.  We then find
{\em any} subgraph of $z^\prime[m]$ that provides this maximum flow and
scale the subgraph so obtained to provide the desired flow.  The cost of
the scaled subgraph is assumed to be the cost of the solution at the end
of each iteration.  We chose the step sizes as follows:
$\alpha_{ij}^{(t)}[m] = \alpha$, $\beta_i^{(t)}[m] = 20 \alpha$, and
$\gamma_{ij}^{(t)}[m]$ was chosen to be large.  The algorithm was run
under two choices of $\alpha$.  We see, from our results, that the value
of $\alpha$ has to be carefully chosen.  Larger values of $\alpha$
generally lead to more oscillatory behavior but faster convergence.

Finally, we considered unicast in lossy wireless networks.  We
conducted simulations where nodes were again placed randomly according
to a uniform distribution over a square region.
The size of
square was set to achieve unit node density.  We considered a network
where transmissions were subject to distance attenuation and Rayleigh
fading, but not interference (owing to scheduling).  So, when node $i$
transmits, the signal-to-noise ratio (SNR) of the signal received at
node $j$ is 
$\gamma d(i,j)^{-2}$,
where $\gamma$ is an exponentially-distributed random variable with unit
mean and $d(i,j)$ is the distance between node $i$ and node $j$.
We assumed that a packet transmitted by node $i$ is successfully
received by node $j$ if the received SNR exceeds $\beta$, i.e.
$\gamma d(i,j)^{-2} \ge \beta$,
where $\beta$ is a threshold that we took to be $1/4$.  If a packet is
not successfully received, then it is completely lost.

We considered five different approaches to wireless unicast;
approaches (\ref{end_retrans})--(\ref{link_retrans}) do not use network
coding, while approaches (\ref{path_coding}) and (\ref{end_coding}) do:
\begin{enumerate}

\item \textbf{End-to-end retransmission:} A path is chosen from source
to sink, and packets are acknowledged by the sink, or destination node.
If the acknowledgment for a packet is not received by the source, the
packet is retransmitted.  This represents the situation where
reliability is provided by a retransmission scheme above the link layer,
e.g.,~by the transport control protocol (TCP) at the transport layer,
and no mechanism for reliability is present at the link layer.
\label{end_retrans}

\item \textbf{End-to-end coding:} A path is chosen from source to sink,
and an end-to-end forward error correction (FEC) code, such as a
Reed-Solomon code, an LT code \cite{lub02}, or a Raptor code
\cite{sho04}, is used to correct for packets lost between source and
sink.  

\item \textbf{Link-by-link retransmission:} A path is chosen from source
to sink, and automatic repeat request (ARQ) is used at the link layer to
request the retransmission of packets lost on every link in the path.
Thus, on every link, packets are acknowledged by the intended receiver
and, if the acknowledgment for a packet is not received by the sender,
the packet is retransmitted.
\label{link_retrans}

\item \textbf{Path coding:} A path is chosen from source to sink, and
every node on the path employs coding to correct for lost packets.  The
most straightforward way of doing this is for each node to use one of
the FEC codes for end-to-end coding, decoding and re-encoding packets it
receives.  The main drawback of such an approach is delay.  Every node
on the path codes and decodes packets in a block.  A way of overcoming
this drawback is to use codes that operate in a more of a
``convolutional'' manner, sending out coded packets formed from packets
received thus far, without decoding.  The random linear coding scheme
from \cite{lmk} is such a code.  A variation, with lower
complexity, is presented in \cite{pfs05}.
\label{path_coding}

\item \textbf{Full coding:} In this case, paths are eschewed altogether.
Problem (\ref{eqn:400}) is solved to find a subgraph,
and the random linear coding scheme from \cite{lmk}
is used.  This represents the limit of
achievability provided that we are restricted from modifying the design
of the physical layer and that we do not exploit the timing of packets
to convey information.
\label{end_coding}

\end{enumerate} 
In all cases
where acknowledgments are sent, acknowledgments are subject to loss in
the same way that packets are and follow the same path.

\begin{figure}
\begin{center}
\includegraphics[height=2in]{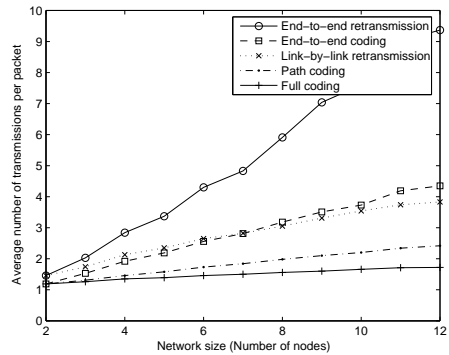}
\end{center}
\caption{Average number of transmissions required per packet using
various wireless unicast approaches in random networks of varying size.
Sources and sinks were chosen randomly according to a uniform
distribution.}
\label{fig:wucostplot}
\end{figure}

The average number of transmissions required per packet using the
various approaches in random networks of varying size is shown in
Figure~\ref{fig:wucostplot}.    
Paths or subgraphs were chosen in each random instance to
minimize the total number of transmissions required, except in the cases
of end-to-end retransmission and end-to-end coding, where they were
chosen to minimize the number of transmissions required by the source
node (the optimization to minimize the total number of transmissions in
these cases cannot be done straightforwardly by a shortest path
algorithm).
We see that, while end-to-end coding and
link-by-link retransmission already represent significant improvements
on end-to-end retransmission, the network coding approaches represent
more significant improvements still.  By a network size of nine nodes,
full coding already improves on link-by-link retransmission by a factor
of two.  Moreover, as the network size grows, the performance of the
various schemes diverges.  Here, we discuss performance simply in terms
of the number of transmissions required per packet; in some cases,
e.g.,~congestion, the performance measure increases super-linearly in
this quantity, and the performance improvement is even greater than that
depicted in Figure~\ref{fig:wucostplot}.  We see, at any rate, that the
use of network coding promises significant improvements, particularly
for large networks.  

\section{Dynamic multicast}
\label{sec:dynamic}

In many applications, membership of the multicast group
changes in time, with nodes joining and leaving the group, rather than
remaining constant for the duration of the connection, as we have thus
far assumed.  
Under these
dynamic conditions, we often cannot simply re-establish the connection
with every membership change because doing so would cause an
unacceptable disruption in the service being delivered
to those nodes remaining in the group.  A good example of an application
where such issues arise is real-time media distribution.
Thus, we desire to find minimum-cost time-varying 
subgraphs that can deliver continuous service to dynamic multicast
groups.

Although our objective is clear, our description of the problem is
currently vague.  Indeed, one of the principal hurdles to tackling the
problem of
dynamic multicast lies in formulating the problem in such a way
that it is suitable for analysis and addresses our objective.
For routed networks, the problem is generally
formulated as the
dynamic Steiner tree problem, which was first proposed in \cite{imw91}.
Under this formulation, the focus is on worst-case behavior and
modifications of the multicast
tree are allowed only when nodes join or leave
the multicast group.  The formulation is adequate, but not compelling;
indeed, there is no compelling reason 
for the restriction on when the multicast tree can be modified.

In our formulation for coded networks,
we draw some inspiration from \cite{imw91},
but we focus on expected behavior rather than worst-case behavior, and
we do not restrict modifications of the multicast subgraph to when nodes
join or leave the multicast tree.
We focus on wireline
networks for simplicity, though our considerations apply 
equally to wireless networks.
We formulate the problem as follows.  

We employ a basic unit of time that
is related to the time that it takes for
changes in the multicast subgraph to settle.  In particular, suppose
that at a given time the multicast subgraph is $z$ and that it is
capable of supporting a multicast connection to
sink nodes $T$.  Then, in one unit
time, we can change the multicast subgraph to $z^\prime$, which is
capable of supporting a multicast connection to 
sink nodes $T^\prime$, without
disrupting the service being delivered to $T \cap T^\prime$ provided
that (componentwise)
$z \ge z^\prime$ or $z \le z^\prime$.  The interpretation of this
assumption is that we allow, in one time unit, only for the subgraph to
increase, meaning that any sink node receiving a particular stream will
continue to receive it (albeit with possible changes in the code,
depending on how the coding is implemented) and therefore facing no
significant disruption to service; or for the subgraph to decrease,
meaning that any sink node receiving a particular stream will be forced to
reduce to a subset of that stream, but one that is sufficient to recover
the source's transmission provided that the sink node is in $T^\prime$, and
therefore again facing no significant disruption to service.  We do not
allow for both operations to take place in a single unit of time (which
would allow for arbitrary changes) because, in that case, sink nodes
may face temporary disruptions to service when decreases to the
multicast subgraph follow too closely to increases.

\begin{figure}
\psfrag{1}{1}
\psfrag{2}{2}
\psfrag{3}{3}
\psfrag{4}{4}
\centering
\includegraphics[scale=1.2]{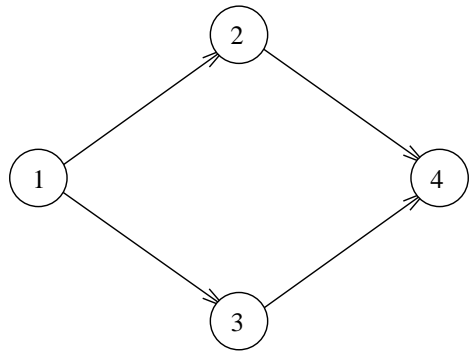}
\caption{A four node network.}
\label{fig:four-node}
\end{figure}

As an example, consider the four node network shown in
Figure~\ref{fig:four-node}.
Suppose that $s = 1$ and that, at a given time, we have $T = \{2,4\}$.
We support a multicast of unit rate with the subgraph 
\[
(z_{12}, z_{13}, z_{24}, z_{34}) = (1,0,1,0).
\]
Now suppose that the group membership changes, and node
2 leaves while node 3 joins, so $T^\prime = \{3,4\}$.  
As a result, we decide that we wish to
change to the subgraph
\[
(z_{12}, z_{13}, z_{24}, z_{34}) = (0,1,0,1).
\]
If we simply make the change na{\"\i}vely in a single time unit, 
then node 4 may face a temporary
disruption to its service as packets on $(2,4)$ stop arriving and before
packets on $(3,4)$ start arriving.  The assumption that we have made on allowed
operations ensures that we must first increase the subgraph to
\[
(z_{12}, z_{13}, z_{24}, z_{34}) = (1,1,1,1),
\]
allow for the change to settle by waiting for one time unit, then
decrease the subgraph to 
\[
(z_{12}, z_{13}, z_{24}, z_{34}) = (0,1,0,1).
\]
With this series of operations, node 4 maintains continuous service
throughout the subgraph change.

We discretize the time axis into time intervals of a single time unit.
We suppose that at the beginning of each time interval, we receive
zero or more
requests from sink nodes that are not currently part of the multicast
group to join and 
zero or more
requests from sink nodes that are currently part of
the multicast group to leave.  We model these join and leave requests as
a discrete stochastic process and make the assumption that, once all the
members of the multicast group leave, the connection is over and remains
in that state forever.  Let $T_m$ denote the sink nodes in the multicast
group at the end of time interval $m$.  Then, we assume that
\begin{equation}
\lim_{m \rightarrow \infty} \Pr(T_m \neq \emptyset | T_0 = T) = 0
\label{eqn:50}
\end{equation}
for any initial multicast group $T$.  A possible, simple model of join 
and leave requests is to model $|T_m|$ as a birth-death process with a
single absorbing state at state 0, and to choose a node uniformly from 
$\mathcal{N}^\prime \setminus T_m$, where $\mathcal{N}^\prime :=
\mathcal{N} \setminus \{s\}$,
 at each birth and from $T_m$ at each death.

Let $z^{(m)}$ be the multicast subgraph at the beginning of time
interval $m$, which, by the assumptions made thus far, means that it
supports a multicast connection to sink nodes $T_{m-1}$.  
Let $V_{m-1}$ and $W_{m-1}$ be the join and leave 
requests that arrive at the end of time interval $m-1$,
respectively.  Hence, $V_{m-1} \subset \mathcal{N}^\prime \setminus T_{m-1}$, 
$W_{m-1} \subset T_{m-1}$, and 
$T_m = (T_{m-1} \setminus W_{m-1}) \cup V_{m-1}$.
We choose $z^{(m+1)}$ from $z^{(m)}$ and $T_m$ using the function
$\mu_m$, so $z^{(m+1)} = \mu_m(z^{(m)}, T_m)$, where $z^{(m+1)}$ must
lie in a particular constraint set $U(z^{(m)}, T_m)$.

To characterize the constraint set $U(z, T)$, 
recall the optimization problem for minimum-cost multicast in
wireline packet networks developed in Section~\ref{sec:wireline}:
\begin{equation}
\begin{split}
& \begin{aligned}
\text{minimize }   & f(z) \\
\text{subject to } 
& z \in Z, \\
& z_{ij} \ge x_{ij}^{(t)} \ge 0,
  \qquad \text{$\forall$ $(i,j) \in A$, $t \in T$} , 
\end{aligned} \\
&\; \sum_{\{j | (i,j) \in A\}} x_{ij}^{(t)}
  - \sum_{\{j | (j,i) \in A\}} x_{ji}^{(t)} 
= \sigma_i^{(t)}, \\
& \qquad \qquad \qquad \qquad \qquad \qquad
\qquad \text{$\forall$ $i \in N$, $t \in T$} , 
\end{split}
\label{eqn:290}
\end{equation}
Therefore,
it follows that we can write
$U(z,T) = U_+(z,T) \cup U_-(z,T)$, where 
\begin{gather*}
U_+(z,T) = \{z^\prime \in Z(T) | z^\prime \ge z\}, \\
U_-(z,T) = \{z^\prime \in Z(T) | z^\prime \le z\},
\end{gather*}
and $Z(T)$ is the feasible set of problem (\ref{eqn:290}) for a given
$T$; i.e.\  if we have the subgraph $z$ at the beginning of a time
interval, and we must go to a subgraph that supports multicast to $T$,
then the allowable subgraphs are those that support multicast to $T$ and
either increase $z$ (those in
$U_+(z,T)$) or decrease $z$ (those in $U_-(z,T)$).

Note that, 
if we have separable constraints,
then $U(z^{(m)}, T_m) \neq \emptyset$ for all $z^{(m)} \in Z$
provided that $Z(T_m) \neq \emptyset$; 
that is, from any feasible subgraph at stage $m$,
it is possible to go to a feasible subgraph
at stage $m+1$ provided that one exists for the
multicast group $T_m$.  But
while this is the case for coded networks, it is not always the case for
routed networks.  Indeed, if multiple multicast trees are being used (as
discussed in \cite{wcj04}, for example), then it is definitely possible
to find ourselves in a state where we cannot achieve multicast 
at stage $m+1$ even though static multicast to $T_m$ 
is possible using multiple multicast trees.  

\begin{figure}
\psfrag{1}{1}
\psfrag{2}{2}
\psfrag{3}{3}
\psfrag{4}{4}
\psfrag{5}{5}
\psfrag{6}{6}
\psfrag{7}{7}
\psfrag{8}{8}
\centering
\includegraphics[scale=1.2]{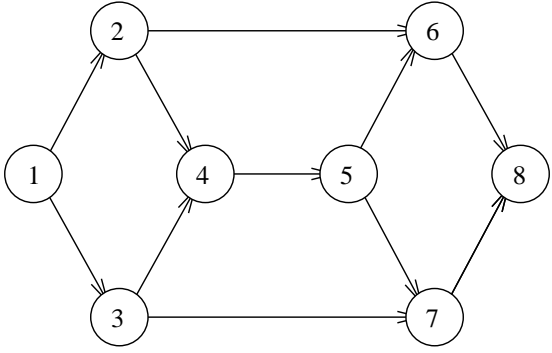}
\caption{A network used for dynamic multicast.}
\label{fig:buttvar}
\end{figure}

As an example of this phenomenon, consider the network
depicted in Figure~\ref{fig:buttvar}.  
Suppose that each arc is of unit capacity,
that $s = 1$, and that, at a given time, we have $T =
\{6, 8\}$.  We support a multicast of rate 2 with the two trees 
$\{(1,3), (3,4), (4,5), (5,6), (5,7), (7,8)\}$ and 
$\{(1,2), (2,6), (6,8)\}$, each carrying unit rate.  Now suppose that
the group membership changes, and node 6 leaves while node 7 joins, so
$T^\prime = \{7, 8\}$.  It is clear that static multicast to $T^\prime$
is possible using multiple multicast trees (we simply
reflect the solution for $T$), but we cannot achieve multicast to
$T^\prime$ by only adding edges to the two existing trees.
Our only recourse at this stage is to abandon the existing trees and
establish new ones, which causes a disruption to the service of node 8,
or to slowly reconfigure the existing trees, which causes a delay before
node 7 is actually joined to the group.

Returning to the problem at hand,
we see that
our objective is to find a policy $\pi = \{\mu_0, \mu_1, \ldots, \}$
that minimizes the cost function
\[
J_\pi(z^{(0)}, T_0)
= \lim_{M \rightarrow \infty} \mathbb{E} \left[
\sum_{m=0}^{M-1} f(z^{(m+1)}) \chi_{2^{\mathcal{N}^\prime} \setminus
\{\emptyset\}}(T_m)
\right] ,
\]
where $\chi_{2^{\mathcal{N}^\prime} \setminus \{\emptyset\}}$ is the characteristic
function for
$2^{\mathcal{N}^\prime} \setminus \{\emptyset\}$ (i.e.\  
$\chi_{2^{\mathcal{N}^\prime} \setminus \{\emptyset\}}(T) = 1$ if $T \neq \emptyset$,
and $\chi_{2^{\mathcal{N}^\prime} \setminus \{\emptyset\}}(T) = 0$ if $T = \emptyset$).

We impose the assumption that
we have separable constraints and that 
$Z(\mathcal{N}^\prime) \neq \emptyset$;
that is, we assume that
there exists a subgraph that supports broadcast.  This
assumption ensures that the constraint set $U(z,T)$ is non-empty for all
$z \in Z$ and $T \subset \mathcal{N}^\prime$. 
Thus, from condition (\ref{eqn:50}), it follows that there exists at
least one policy $\pi$ (namely, one that uses some fixed
$z \in Z(\mathcal{N}^\prime)$ until the multicast group is empty) such that
$J_\pi(z^{(0)}, T_0) < \infty$.

It is now not difficult to see that we are dealing with an undiscounted,
infinite-horizon dynamic programming problem (see, for example,
\cite[Chapter 3]{ber01b}), and we can apply the theory
developed for such problems to our problem.  
So doing, 
we first note that the optimal cost function $J^* := \min_\pi J_\pi$
satisfies Bellman's equation; namely, we have
\begin{equation*}
J^*(z, T) 
=
\min_{u \in U(z, T)} \left\{ f(u)
+ \mathbb{E} [ J^*(u, (T \setminus V) \cup W) ]
\right\} 
\end{equation*}
if $T \neq \emptyset$, and $J^*(z,T) = 0$ if $T = \emptyset$.
Moreover, the optimal cost is achieved by the stationary policy $\pi =
\{\mu, \mu, \ldots\}$, where $\mu$ is given by
\begin{equation}
\mu(z, T) 
=
\argmin_{u \in U(z, T)} \left\{ f(u)
+ \mathbb{E} [ J^*(u, (T \setminus V) \cup W) ]
\right\} 
\label{eqn:300}
\end{equation}
if $T \neq \emptyset$, and $\mu(z,T) = 0$ if $T = \emptyset$.

The fact that the optimal cost can be achieved by a stationary policy
limits the space in which we need to search for optimal policies
significantly, but we are still left with the difficulty that the state
space is uncountably large; it is the space of all possible pairs $(z, T)$,
which is $Z \times 2^{\mathcal{N}^\prime}$.
The size of the state space more or less
eliminates the possibility of using
techniques such as value iteration to obtain $J^*$.

On the other hand, given $J^*$, it does not seem at all implausible that
we can compute the optimal decision at the beginning of each time
interval using (\ref{eqn:300}).  Indeed, the constraint set is the union
of two polyhedra, which can be handled by optimizing over each
separately, and, although the objective function may not necessarily be
convex even if $f$ is convex owing to the term 
$\mathbb{E} [ J^*(u, (T \setminus V) \cup W) ]$, 
we are, at any rate, unable to obtain $J^*$ precisely on account of
the large state space, and can restrict our attention to approximations
that make problem (\ref{eqn:300}) tractable.

For dynamic programming problems, there are many approximations that
have been developed to cope with large state spaces (see, for example,
\cite[Section 2.3.3]{ber01b}).  In particular, we can
approximate $J^*(z, T)$ by $\tilde{J}(z, T, r)$, where $\tilde{J}(z, T,
r)$ is of some fixed form, and $r$ is a parameter vector that is
determined by some form of optimization, which can be performed offline
if the graph $\mathcal{G}$ is static.
Depending upon the approximation that is used, we may even be able to
solve problem (\ref{eqn:300}) using the decentralized algorithms
described in Section~\ref{sec:wireline} (or simple modifications
thereof).
The specific approximations $\tilde{J}(z, T, r)$
that we can use and their performance
are beyond the scope of this paper.

\section{Conclusion}
\label{sec:conclusion}

Routing is certainly a satisfactory way to operate packet
networks.  It clearly works, but 
it is not clear that it should be used for all types of
networks.  As we have mentioned, application-layer overlay networks and
multi-hop wireless networks are two types of networks where coding is a
definite alternative.

To actually use coding, however, we must apply to coding the same
considerations that we normally apply to routing.  This paper did
exactly that:  We took the cost consideration from routed packet
networks and applied it to coded packet networks.  More specifically, we
considered 
the problem of finding minimum-cost subgraphs
to support multicast connections over coded
packet networks---both wireline and wireless.  
As we saw, this problem is
effectively decoupled from the coding problem:  To establish
minimum-cost multicast connections, we can first
determine the rate to inject coded packets on each arc, then determine 
the contents of those packets.  

Our work therefore brings coded packet networks one step closer to
realization.  But, to actually see that happen, much work remains to be
done.  For example, designing protocols around our algorithms is a clear
task, as is designing protocols to implement coding schemes.  
In addition, there are some important issues coming directly
from this paper
that require further exploration.  Some of these relate to the
decentralized algorithms, e.g.,\  
their stability under changing
conditions (e.g.,\  changing arc costs, changing graph topology), 
their speeds of convergence, their demands on computation and
message-exchange, and their behavior under asynchronism.
Another topic to explore is specific
approximation methods for use in our formulation of dynamic multicast.

On a broader level, we could design other algorithms using the flow
formulations given in this paper (see \cite{hov05, xiy05}).  And we
could give more thought to the cost functions themselves.  Where do they
come from?  Do cost functions for routed packet networks make sense for
coded ones?  If a coded packet network is priced, how should the pricing
be done?  And how should the resultant cost be shared among the members
of the multicast group?

In short, we believe that realizing coded packet networks is a
worthwhile goal, and we see our work as an integral step toward this
goal.
Much promising work, requiring various expertise, remains.

\section*{Acknowledgments}

The authors
would like to thank R. Srikant for helpful
discussions and suggestions and Hyunjoo Lee for her work on the
simulation software.

\appendices

\section{}
\label{app:1}

We wish to solve the following problem.
\begin{equation*}
\begin{aligned}
\text{minimize } & \sum_{t \in T} (v^{(t)} - u^{(t)})^2 \\
\text{subject to } & v \in P_{ij} ,
\end{aligned}
\end{equation*}
where $P_{ij}$ is the $|T|$-dimensional simplex
\begin{equation*}
P_{ij} = 
\left\{v \left| 
\sum_{t \in T} v^{(t)} = a_{ij},\,
v \ge 0 \right.\right\} .
\end{equation*}
First, since the objective function and the constraint set $P_{ij}$ are
both convex, it is straightforward to establish that a necessary and
sufficient condition for global optimality of $\hat{v}^{(t)}$ in $P_{ij}$ 
is
\begin{equation}
\hat{v}^{(t)} > 0 \Rightarrow 
(u^{(t)} - \hat{v}^{(t)}) \ge (u^{(r)} - \hat{v}^{(r)}),
\qquad \text{$\forall$ $r \in T$}
\label{eqn:100}
\end{equation}
(see, for example, \cite[Section 2.1]{ber95}).
Suppose we index the elements of $T$ such
that $u^{(t_1)} \ge u^{(t_2)} \ge \ldots \ge u^{(t_{|T|})}$.
We then note that there must be an index $k$ in the set 
$\{1, \ldots, |T|\}$ such that $v^{(t_l)} > 0$
for $l = 1, \ldots, k$ and $v^{(t_l)} = 0$ for $l > k+1$, 
for, if not, then a feasible solution with lower cost can be obtained by
swapping around components of the vector.
Therefore, condition (\ref{eqn:100}) implies that there must exist some
$d$ such that ${\hat{v}^{(t)}} = u^{(t)} + d$ for all $t \in
\{t_1, \ldots, t_k\}$ and that $d \le -u^{(t)}$ for all $t \in
\{t_{k+1}, \ldots, t_{|T|}\}$, which is equivalent to
$d \le -u^{(t_{k+1})}$.  
Since ${\hat{v}^{(t)}}$ is in the simplex $P_{ij}$, it follows that
\begin{equation*}
kd + \sum_{t = 1}^{t_k} u^{(t)} = a_{ij},
\end{equation*}
which gives
\begin{equation*}
d = \frac{1}{k}\left(a_{ij} - \sum_{t = 1}^{t_k} u^{(t)}\right) .
\end{equation*}
By taking $k = \hat{k}$, where $\hat{k}$ is the smallest $k$ such that
\begin{equation*}
\frac{1}{k}\left(a_{ij} - \sum_{r=1}^{t_k} u^{(r)} \right) 
\le -u^{(t_{k+1})},
\end{equation*}
(or, if no such $k$ exists, then $\hat{k} = |T|$), we see that we have
\begin{equation*}
\frac{1}{\hat{k}-1}\left(a_{ij} - \sum_{t=1}^{t_{k-1}}u^{(t)}\right)
> -u^{(t_k)},
\end{equation*}
which can be rearranged to give
\begin{equation*}
d = \frac{1}{\hat{k}}\left(a_{ij} - \sum_{t=1}^{t_k}u^{(t)}\right)
> -u^{(t_k)} .
\end{equation*}
Hence, if $v^{(t)}$ is given by
\begin{equation}
v^{(t)} = 
\begin{cases}
u^{(t)} + \frac{a_{ij} 
- \sum_{r=1}^{t_{\hat{k}}} u^{(r)}}{\hat{k}}
  & \text{if $t \in \{t_{1}, \ldots, t_{\hat{k}}\}$}, \\
0 & \text{otherwise} ,
\label{eqn:110}
\end{cases}
\end{equation}
then $v^{(t)}$ is feasible and we see that the optimality condition
(\ref{eqn:100}) is satisfied.  Note that, since $d \le -u^{(t_{k+1})}$,
equation (\ref{eqn:110}) can
also be written as
\begin{equation}
v^{(t)} = \max\left(0, u^{(t)} + \frac{1}{\hat{k}}\left(a_{ij} -
\sum_{r=1}^{t_{\hat{k}}} u^{(r)}\right) \right) .
\label{eqn:115}
\end{equation}

We now turn to showing that 
any accumulation point of the sequence of primal iterates
$\{x[n]\}$ given by (\ref{eqn:17}) is an optimal solution the primal
problem (\ref{eqn:2}).  Suppose that the dual feasible solution that the
subgradient method converges to is $\bar{p}$. 
Then there exists some $m$ such that for $n \ge m$
\begin{equation*}
p_{ij}^{(t)}[n+1] = p_{ij}^{(t)}[n] + \theta[n]x_{ij}^{(t)}[n]
+ c_{ij}[n]
\end{equation*}
for all $(i,j) \in \mathcal{A}$ and $t \in T$ such that 
$\bar{p}_{ij}^{(t)} > 0$.  
Therefore, if $\bar{p}_{ij}^{(t)} > 0$, then for $n > m$ we have
\begin{equation}
\begin{split}
\tilde{x}_{ij}^{(t)}[n] &= \sum_{l=1}^m\mu_l[n]x_{ij}^{(t)}[l] 
    + \sum_{l=m+1}^n\mu_l[n]x_{ij}^{(t)}[l] \\
  &= \sum_{l=1}^m\mu_l[n]x_{ij}^{(t)}[l] \\
  &\phantom{=}
    + \sum_{l=m+1}^n\frac{\mu_l[n]}{\theta[n]}
      (p_{ij}^{(t)}[n+1] - p_{ij}^{(t)}[n] - d_{ij}[n]) \\
  &= \sum_{l=1}^m\mu_l[n]x_{ij}^{(t)}[l] 
    + \sum_{l=m+1}^n\gamma_{ln}(p_{ij}^{(t)}[n+1] - p_{ij}^{(t)}[n]) \\
  &\phantom{=}
    - \sum_{l=m+1}^n\gamma_{ln}d_{ij}[n] .
\end{split}
\label{eqn:120}
\end{equation}
Otherwise, if $\bar{p}_{ij}^{(t)} = 0$, then from equation
(\ref{eqn:115}), we have
\begin{equation*}
p_{ij}^{(t)}[n+1] \ge p_{ij}^{(t)}[n] + \theta[n]x_{ij}^{(t)}[n]
+ c_{ij}[n] ,
\end{equation*}
so
\begin{equation}
\begin{split}
\tilde{x}_{ij}^{(t)}[n] &\le \sum_{l=1}^m\mu_l[n]x_{ij}^{(t)}[l] 
    + \sum_{l=m+1}^n\gamma_{ln}(p_{ij}^{(t)}[n+1] - p_{ij}^{(t)}[n]) \\
&\phantom{\le}
    - \sum_{l=m+1}^n\gamma_{ln}c_{ij}[n] .
\label{eqn:130}
\end{split}
\end{equation}

It is straightforward to see that the sequence of iterates
$\{\tilde{x}[n]\}$ is primal feasible, and that we obtain a primal
feasible sequence $\{z[n]\}$ by setting
$z_{ij}[n] := \max_{t \in T} \tilde{x}_{ij}^{(t)}[n]$.
Sherali and Choi \cite{shc96} showed that, if the required conditions on
the step sizes $\{\theta[n]\}$ and convex combination weights
$\{\mu_l[n]\}$ are satisfied, then 
\begin{equation*}
\sum_{l=1}^m\mu_l[n]x_{ij}^{(t)}[l]
    + \sum_{l=m+1}^n\gamma_{ln}(p_{ij}^{(t)}[n+1] - p_{ij}^{(t)}[n])
\rightarrow 0
\end{equation*}
as $k \rightarrow \infty$;
hence we see from equations (\ref{eqn:120}) and (\ref{eqn:130}) that, 
for $k$ sufficiently large, 
\begin{equation*}
z_{ij}[n] = - \sum_{l=m+1}^n\gamma_{ln}c_{ij}[n] 
\end{equation*}
and, therefore, that 
complementary slackness with $\bar{p}$
holds in the limit of any convergent subsequence
of $\{\tilde{x}[n]\}$.

\section{Proof of Proposition~\ref{prop:stability}}
\label{app:stability}

We prove the stability of the primal-dual algorithm by using the theory
of Lyapunov stability (see, for example, \cite[Section 3.10]{sri04}).
This proof is based on the proof of Theorem 3.7 of \cite{sri04}.

The Lagrangian for problem (\ref{eqn:30}) is as follows:
\begin{multline}
L(x,p,\lambda) 
= U(x) \\
- 
\sum_{t \in T} \left\{ \sum_{i \in N}
p_i^{(t)} \left( \sum_{\{j | (i,j) \in A\}} x_{ij}^{(t)} 
  - \sum_{\{j | (j,i) \in A\}} x_{ji}^{(t)} 
\right.\right. \\ \left.\left.
- \sigma_i^{(t)} \right)
- \sum_{(i,j) \in A} \lambda_{ij}^{(t)} x_{ij}^{(t)} \right\} .
\label{eqn:18}
\end{multline}
The function $U$ is strictly concave since $f_{ij}$ is a monotonically
increasing, strictly convex function and $z_{ij}^\prime$ is a strictly
convex function of $x_{ij}$, so
there exists
a unique minimizing solution for problem (\ref{eqn:30}), 
say $\hat{x}$,
and Lagrange multipliers, say $\hat{p}$ and $\hat{\lambda}$, which
satisfy the following Karush-Kuhn-Tucker conditions.
\begin{gather}
\allowdisplaybreaks
\begin{gathered}
\frac{\partial L(\hat{x},\hat{p},\hat{\lambda})}
{\partial x_{ij}^{(t)}} = \left(\frac{\partial U(\hat{x})}
{\partial x_{ij}^{(t)}} - \left(\hat{p}_i^{(t)} - \hat{p}_j^{(t)}\right) 
+ \hat{\lambda}_{ij}^{(t)}\right) = 0 , \\
\qquad \qquad \qquad \qquad \qquad \qquad
\qquad \text{$\forall$ $(i,j) \in A$, $t \in T$} , 
\end{gathered} 
\label{eqn:kkt1} \\
\begin{gathered}
\sum_{\{j | (i,j) \in A\}} \hat{x}_{ij}^{(t)}
  - \sum_{\{j | (j,i) \in A\}} \hat{x}_{ji}^{(t)} 
= \sigma_i^{(t)}, \\
\qquad \qquad \qquad \qquad \qquad \qquad 
\qquad \text{$\forall$ $i \in N$, $t \in T$} , 
\end{gathered} \\
\hat{x}_{ij}^{(t)} \geq 0 \qquad 
\text{$\forall$ $(i,j) \in A$, $t \in T$} , \\
\hat{\lambda}_{ij}^{(t)} \geq 0 \qquad 
\text{$\forall$ $(i,j) \in A$, $t \in T$} , \\
\hat{\lambda}_{ij}^{(t)} \hat{x}_{ij} = 0 \qquad
\text{$\forall$ $(i,j) \in A$, $t \in T$} . 
\label{eqn:kkt4}
\end{gather}

From Equation (\ref{eqn:18}), it can be verified that $(\hat{x},
\hat{p},\hat{\lambda})$ is an equilibrium point of the
primal-dual algorithm.  We now prove that this point is globally,
asymptotically stable.

Consider the following function as a candidate for the
Lyapunov function:
\begin{multline*}
V(x,p,\lambda) \\
= \sum_{t \in T}
\left\{
\sum_{(i,j) \in A} \left( \int_{\hat{x}_{ij}^{(t)}}^{x_{ij}^{(t)}}
\frac{1}{k_{ij}^{(t)}(\sigma)}(\sigma - \hat{x}_{ij}^{(t)}) d\sigma
\right.\right.  \\
\left.\left.
+ \int_{\hat{\lambda}_{ij}^{(t)}}^{\lambda_{ij}^{(t)}}
\frac{1}{m_{ij}^{(t)}(\gamma)}(\gamma - \hat{\lambda}_{ij}^{(t)})
d\gamma \right)
\right. \\
\left.
+ \sum_{i \in N} \int_{\hat{p}_i^{(t)}}^{p_i^{(t)}}
\frac{1}{h_i^{(t)}(\beta)}(\beta - \hat{p}_i^{(t)}) d\beta
\right\} .
\end{multline*}
Note that $V(\hat{{x}},\hat{{p}},\hat{\lambda}) = 0$. Since,
$k_{ij}^{(t)}(\sigma) > 0$, if $x_{ij}^{(t)} \neq \hat{x}_{ij}^{(t)}$, 
we have
$\int_{\hat{x_{ij}^{(t)}}}^{x_{ij}^{(t)}}\frac{1}{k_{ij}^{(t)}
(\sigma)}(\sigma - \hat{x}_{ij}^{(t)})d\sigma >0$. 
This argument can be extended to the other
  terms as well.
Thus, whenever $({x}, {p},{\lambda}) \neq (\hat{x}
,\hat{{p}},\hat{{\lambda}})$, we have $V({x},{p},{\lambda}) > 0$.

Now,
\begin{multline*}
\dot{V}
= \sum_{t \in T}
\left\{
\sum_{(i,j) \in A} \left[ 
\left(-x_{ij}^{(t)}\right)^+_{\lambda_{ij}^{(t)}}
(\lambda_{ij}^{(t)} - \hat{\lambda}_{ij}^{(t)})
\right.\right. \\ \left.\left.
+ \left(\frac{\partial U(x)}{\partial x_{ij}^{(t)}}
- q_{ij}^{(t)} + \lambda_{ij}^{(t)} \right) 
\cdot (x_{ij}^{(t)} - \hat{x}_{ij}^{(t)})
\right]
\right. \\
\left.
+ \sum_{i \in N}
(y_i^{(t)} - \sigma_i^{(t)})
(p_i^{(t)} - \hat{p}_i^{(t)}) 
\right\} .
\end{multline*}
Note that
\begin{equation*}
\left(-x_{ij}^{(t)}\right)^+_{\lambda_{ij}^{(t)}}
(\lambda_{ij}^{(t)} - \hat{\lambda}_{ij}^{(t)})
\le 
-x_{ij}^{(t)}
(\lambda_{ij}^{(t)} - \hat{\lambda}_{ij}^{(t)}) ,
\end{equation*}
since the inequality is an equality if either $x_{ij}^{(t)} \le 0$ or 
$\lambda_{ij}^{(t)} \ge 0$; and, in the case when $x_{ij}^{(t)} > 0$ and
$\lambda_{ij}^{(t)} < 0$, we have
$(-x_{ij}^{(t)})^+_{\lambda_{ij}^{(t)}} = 0$ and, since
$\hat{\lambda}_{ij}^{(t)} \ge 0$, 
$-x_{ij}^{(t)}(\lambda_{ij}^{(t)} - \hat{\lambda}_{ij}^{(t)}) \ge 0$.
Therefore, 
\begin{equation*}
\begin{split}
\dot{V}
&\le \sum_{t \in T}
\left\{
\sum_{(i,j) \in A} \left[ 
-x_{ij}^{(t)}
(\lambda_{ij}^{(t)} - \hat{\lambda}_{ij}^{(t)})
\right.\right. \\ 
& \qquad \left.\left.+
\left(\frac{\partial U(x)}{\partial x_{ij}^{(t)}}
- q_{ij}^{(t)} + \lambda_{ij}^{(t)} \right) 
\cdot (x_{ij}^{(t)} - \hat{x}_{ij}^{(t)})
\right]
\right. \\
& \phantom{= \sum_{t \in T} \{}
 \left.
+ \sum_{i \in N}
(y_i^{(t)} - \sigma_i^{(t)})
(p_i^{(t)} - \hat{p}_i^{(t)}) 
\right\} \\
&= (\hat{q} - q)^\prime(x - \hat{x})
+ (\hat{p} - p)^\prime(y - \hat{y}) \\
&\phantom{=}
+ \sum_{t \in T}
\left\{
\sum_{(i,j) \in A} \left[ 
-\hat{x}_{ij}^{(t)}
(\lambda_{ij}^{(t)} - \hat{\lambda}_{ij}^{(t)})
\right.\right. \\ 
& \qquad \left.\left.+
\left(\frac{\partial U(x)}{\partial x_{ij}^{(t)}}
- \hat{q}_{ij}^{(t)} + \hat{\lambda}_{ij}^{(t)} \right) 
\cdot (x_{ij}^{(t)} - \hat{x}_{ij}^{(t)})
\right]
\right. \\
& \phantom{= \sum_{t \in T} \{}
 \left.
+ \sum_{i \in N}
(\hat{y}_i^{(t)} - \sigma_i^{(t)})
(p_i^{(t)} - \hat{p}_i^{(t)}) 
\right\} \\
&= (\bigtriangledown U(x) - \bigtriangledown U(\hat{x}))^\prime
(x - \hat{x}) - \lambda^\prime \hat{x} ,
\end{split}
\end{equation*}
where the last line follows from Karush-Kuhn-Tucker conditions
(\ref{eqn:kkt1})--(\ref{eqn:kkt4}) and the fact that
\begin{equation*}
\begin{split}
p^\prime y
&= \sum_{t \in T}
\sum_{i \in \mathcal{N}}
p_{i}^{(t)} 
\left(\sum_{\{j | (i,j) \in \mathcal{A}\}} \hat{x}_{ij}^{(t)}
  - \sum_{\{j | (j,i) \in \mathcal{A}\}} \hat{x}_{ji}^{(t)} \right) \\
&= \sum_{t \in T}
\sum_{(i,j) \in \mathcal{A}}
x_{ij}^{(t)} (p_i^{(t)} - p_j^{(t)}) = q^\prime x .
\end{split}
\end{equation*}
Thus, owing to the strict concavity of $U(x)$, we have
$\dot{V} \le -\lambda^\prime \hat{x}$,
with equality if and only if $x = \hat{x}$.
So it follows that $\dot{V} \le 0$ for all $\lambda \ge 0$, since
$\hat{x} \ge 0$.

If the initial choice of $\lambda$ is such that $\lambda(0) \geq 0$, we
see from the primal-dual algorithm that $\lambda(\tau) \geq 0$. 
This is
true since $\dot{\lambda} \geq 0$ whenever $\lambda \leq 0$. Thus,
it follows by the theory of Lyapunov stability that the algorithm is
indeed
globally, asymptotically stable.

\section{Proof of Proposition~\ref{prop:10}}
\label{app:prop_10}

Suppose $(x, z)$ is a feasible solution to problem (\ref{eqn:422}).
Then, for all $(i,j) \in \mathcal{A}^\prime$ and $t \in T$,
\[
\begin{split}
\sum_{m = m(i,j)}^{M_i} z_{iJ_m^{(i)}}
&\ge \sum_{m = m(i,j)}^{M_i} \sum_{k \in J_m^{(i)}}
x_{iJ_m^{(i)}k}^{(t)} \\
&= \sum_{k \in J_{M_i}^{(i)}} \sum_{m = \max(m(i,j), m(i,k))}^{M_i}
x_{iJ_m^{(i)}k}^{(t)} \\
&\ge \sum_{k \in J_{M_i}^{(i)} \setminus J^{(i)}_{m(i,j)-1}}
\sum_{m = \max(m(i,j), m(i,k))}^{M_i}
x_{iJ_m^{(i)}k}^{(t)} \\
&= \sum_{k \in J_{M_i}^{(i)} \setminus J^{(i)}_{m(i,j)-1}}
\sum_{m = m(i,k)}^{M_i}
x_{iJ_m^{(i)}k}^{(t)} \\
&= \sum_{k \in J_{M_i}^{(i)} \setminus J^{(i)}_{m(i,j)-1}}
\hat{x}_{ik}^{(t)} .
\end{split}
\]
Hence $(\hat{x},z)$ is a feasible solution of problem (\ref{eqn:450})
with the same cost.

Now suppose $(\hat{x}, z)$ is an optimal solution of problem
(\ref{eqn:450}).  
Since $f_{iJ_1^{(i)}}(\zeta) < f_{iJ_2^{(i)}}(\zeta) < \cdots
< f_{iJ_{M_i}^{(i)}}(\zeta)$ for all $\zeta \ge 0$ and
$i \in \mathcal{N}$
by assumption, it
follows that, for all $i \in \mathcal{N}$, the sequence
$z_{iJ_1^{(i)}}, z_{iJ_2^{(i)}}, \ldots, z_{iJ_{M_i}^{(i)}}$ is given
recursively, starting from $m = M_i$, by
\[
z_{iJ_m^{(i)}} = \max_{t \in T}
\left\{
\sum_{k \in J_{M_i}^{(i)} \setminus J_{m-1}^{(i)}} \hat{x}_{ik}^{(t)}
- \sum_{l=m+1}^{M_i} z_{iJ_l^{(i)}}
\right\} .
\]
Hence $z_{iJ_m^{(i)}} \ge 0$ for all $i \in \mathcal{N}$ and $m = 1, 2,
\ldots, M_i$.
We then set, starting from $m = M_i$ and $j \in J_{M_i}^{(i)}$,
\begin{multline*}
x_{iJ_m^{(i)}j}^{(t)} 
:= \min\left(
\hat{x}_{ij}^{(t)} - \sum_{l = m+1}^{M_i} x_{iJ_l^{(i)}j} ,
z_{iJ_m^{(i)}} 
\right. \\
\quad\left.
- \sum_{k \in J_{M_i}^{(i)} \setminus J_{m(i,j)}^{(i)}}
x_{iJ_m^{(i)}k}^{(t)}
\right) .
\end{multline*}
It is now difficult to see that $(x,z)$ is a feasible solution of
problem (\ref{eqn:422}) with the same cost.

Therefore, the optimal costs of problems (\ref{eqn:422}) and
(\ref{eqn:450}) are the same and, since the objective functions for the
two problems are the same, $z$ is part of an optimal solution for
problem (\ref{eqn:422}) if and only if it is part of an optimal solution
for problem (\ref{eqn:450}).

\bibliographystyle{IEEEtran}
\bibliography{IEEEabrv,inform_theory}

\end{document}